\documentclass[12pt,preprint]{aastex}

\shorttitle{Supernova Yield Uncertainties}
\shortauthors{Young et al.}

\newcommand{\sol}{$M_\odot$\,}
\def \nuc#1#2{\relax\ifmmode{}^{#1}{\protect\text{#2}}\else${}^{#1}$#2\fi}

\begin{document}

\title{Uncertainties in Supernova Yields I: 1D Explosions}

\author{Patrick A. Young\altaffilmark{1,2}, Chris L. Fryer\altaffilmark{1,3}}
\altaffiltext{1}{Theoretical Astrophysics, Los Alamos National Laboratories, Los Alamos, NM 87545}
\altaffiltext{2}{Steward Observatory, University of Arizona, 
 Tucson AZ 85721}
\altaffiltext{3}{Physics Dept., University of Arizona, 
 Tucson AZ 85721}

\email{payoung@lanl.gov, fryer@lanl.gov}

\begin{abstract}

Theoretical nucleosynthetic yields from supernovae are sensitive to
both the details of the progenitor star and the explosion
calculation. We attempt to comprehensively identify the sources of
uncertainties in these yields. In this paper we concentrate on the
variations in yields from a single progenitor arising from common
1-dimensional methods of approximating a supernova
explosion. Subsequent papers will examine 3-dimensional effects in the
explosion and the progenitor, and trends in mass and composition. For
the 1-dimensional explosions we find that both elemental and isotopic
yields for Si and heavier elements are a sensitive function of
explosion energy. Also, piston-driven and thermal bomb type explosions
have different yields for the same explosion energy. Yields derived
from 1-dimensional explosions are non-unique.

\end{abstract}


\section{INTRODUCTION}

The amount of observational data on chemical abundances is growing at
an enormous rate. The Sloan Digital Sky Survey has produced spectra
for $> 10^5$ galaxies at low redshifts. Information is also becoming
available for smaller numbers of galaxies to redshifts $> 6$ with the
current generation of large telescopes \citep{berg06}. Observations at
high to moderate redshift show a gradual build-up in metallicity
\citep{erb06}, from which conclusions about the star formation
histories of galaxies are drawn. Comparison of metallicities between
elliptical and spiral galaxies, members of groups and clusters, and
abundance gradients within individual galaxies are used to probe their
formation, interaction history, and evolution. Large telescopes are
providing detailed abundances for Milky Way halo stars and dwarf
galaxies and globular clusters in the local group. The dwarfs and
globular clusters show a range of nearly a dex in [$\alpha$/Fe] at a
given [Fe/H] \citep{pvi05}. Several ultra-metal-poor stars are now
known with [Fe/H] $< -4$ \citep{aoki06}. Each has unique abundance
anomalies that suggest they have been enriched by only one or a few
stars.

Chemical evolution over a large fraction of the age of the universe is
becoming a quantitative field of study. The conclusions that can be
drawn from this data, however, are only as good as our understanding
of how chemical elements are produced in stars. The best currently
practical approach to producing theoretical yields would be to take a
set of progenitor stars and explode them in 1-dimension with a range
of free parameters such as explosion energy or mass cut. Then a linear
combination of the models, weighted by an initial mass function (IMF)
would be chosen which produces the desired abundance pattern, i.e
solar, and these would then form a table of yields with mass and
metallicity.  In fact, the common approach is to take a single set of
yields without any exploration of parameter space.
\citep{gar02}\footnote{The common standard for yields is
\citet{ww95}. This compilation only presents yields for a single
explosion energy, except at progenitor masses of 30 - 40 \sol.}.  In
either scenario, these yields are fit, not predicted.  The local group
results indicate that very different enrichment histories with very
different abundance ratios can lead to the same {\it total} enrichment
in metallicity. Exploring these pathways require predictive yields.
Identifying the star or stars that enriched the ultra-metal-poor halo
stars even more clearly requires predictive yields with a unique
correspondence to a progenitor.

Unfortunately, it is not presently possible routinely (or perhaps at
all) to produce a truly predictive yield for an ensemble of stars. The
nucleosynthesis is sensitive to a number of properties of the
progenitor and explosion models, both physical and numerical. The
structure of progenitors integrated over the evolution changes
dramatically if hydrodynamic processes are included
\citep{ymaf05}. The evolution also depends upon rotation and mass loss
\citep[i.e.][]{mm00}. The dynamics of the progenitor before and during
collapse will also leave an imprint on the explosion. Hydrodynamic
motions can result in asymmetries in the shell burning regions of
$>10\%$ at the onset of collapse \citep{ma06}. The integrated effects
can be incorporated into a stellar evolution code after an analytic
framework is derived from examining simulations, but the late stage
dynamics must be simulated directly with a computationally expensive
multi-dimensional hydrodynamics code. The composition and mass of the
progenitor of course also play a role.

The idea that the convection above the proto-neutron star plays an
important role in understanding supernova energetics is gradually
becoming accepted \citep{her94,fw02,bm06,bur06}.  If so, understanding
the explosion and obtaining accurate explosion energetics will require
3-dimensional calculations.  We are far from modeling all of the
physics, including the convection, with enough detail in 2-dimensions,
let alone 3, to accurately estimate explosion energies.  2-dimensional
simulations assuming a 90$^{\circ}$ wedge geometry \citep{bur03} have
obtained different results when modeled using a 180$^{\circ}$ domain.
(Janka - pvt. communication).  The convective instabilities depend
sensitively on the equation of state, resolution and, probably,
implementation of the gravitational force \citep{fk06,fry06}.  The
convective instabilities also depend on the effects of rotation and
asymmetries in the collapsing core \citep{sys94,fh00,kys03,fw04}.

In addition to being important for our understanding of the explosive
engine, multi-dimensional simulations are required to understand the
outward mixing of heavy elements and, ultimately, the amount of this
material that is ejected in a supernova \citep{Kif00,hun03} This
effect is enhanced by explosion asymmetries
\citep{Nag98,Kif03,hun03,hun05}.  Multi-dimensional studies of the
explosion are just now beginning to put in the relevant physics, and
these new studies will no doubt bring new surprises in supernovae
explosions.  Detailed multi-dimensional models are required both to
understand the supernova engine and the resultant explosion and, at
this point in time, much more work must be done before we can predict
either the explosive energies or the yields from supernovae.  For
massive stars, explosion mechanisms beyond the standard supernova
engine \citep[e.g.][the collapsar model]{woo93}, will also play a role
in nucleosynthetic yields.

Clearly, progress in the fields of chemical evolution, stellar
populations, and galaxy assembly is not going to be put on hold while
all the outstanding issues in supernovae and nucleosynthesis are
resolved. We can, however, attempt to quantify the uncertainties
arising from various assumptions in the yield calculations. In this
series of papers we will identify sources of uncertainty from each
aspect of the calculations so that we may include error bars in a new
generation of integrated yields. In this first paper we will examine
changes in yields for a single progenitor arising from various methods
of performing 1D explosion calculations. Even though we are able to
produce a weak explosion of a 23 \sol star in 3D, the simulation
requires approximately seven months of computer time on a moderately
sized cluster. Doing multi-D explosions for even a sparse sampling of
a population will not be feasible for some time, so we still must rely
on 1D models for integrated yields. In the second paper we will
examine a full 3D collapse and explosion of the same progenitor and
compare it to the 1D explosions. Paper three will study trends with
changes in progenitor mass, composition and evolution code physics. A
final paper will compare our initially spherically symmetric 3D
explosion with one with realistic progenitor dynamics as initial
conditions.

\section{PROGENITOR AND EXPLOSION CALCULATIONS}

\subsection{Progenitor}

\begin{figure}
\figurenum{1}
\plotone{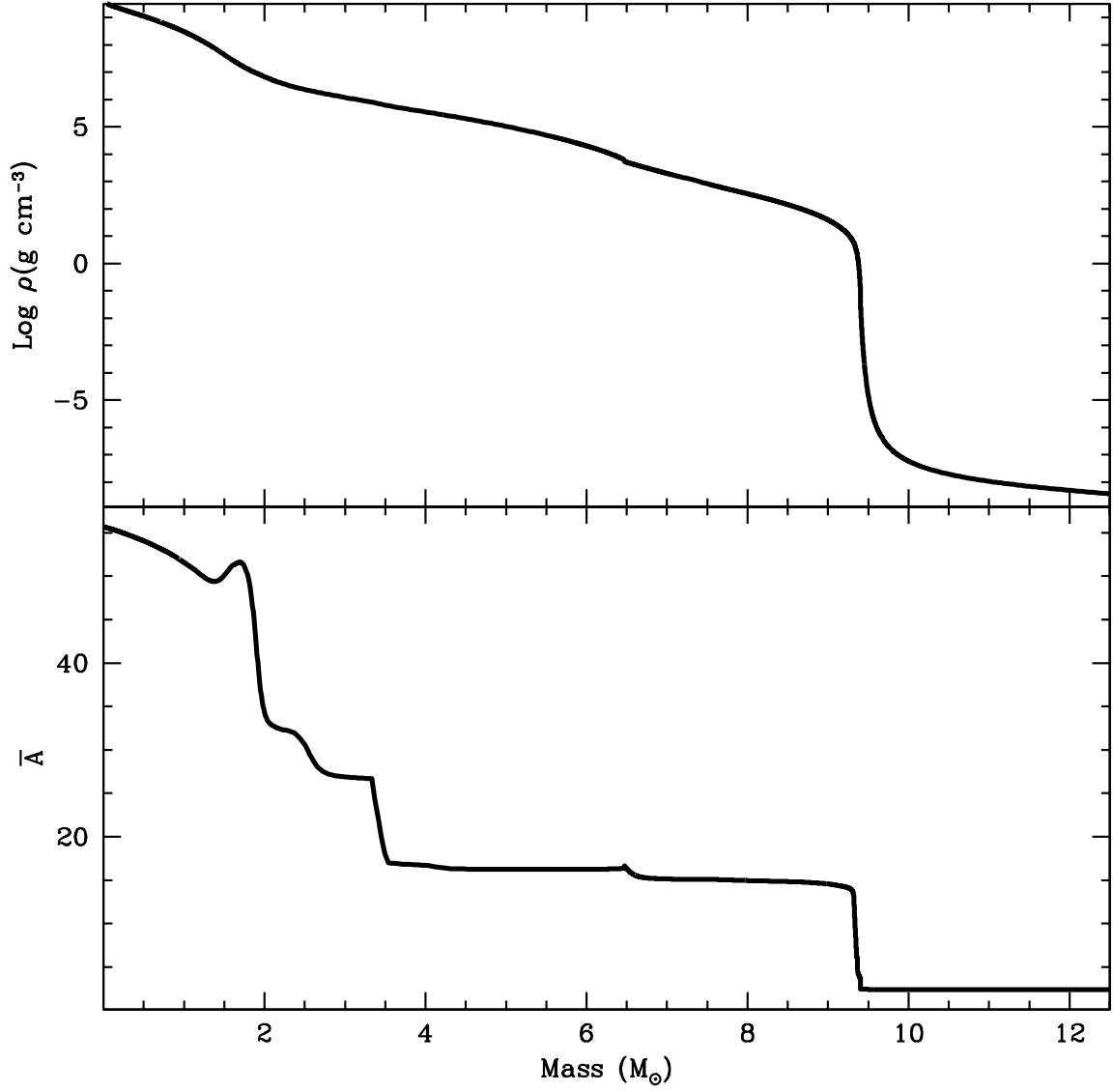}
\caption{Mean atomic weight \={A} (top) and density (bottom) vs. mass
coordinate for the progenitor model. \label{fig1} }
\end{figure}

We use a 23 \sol progenitor of \citet{gs98} solar composition, aspects
of which we have examined in \citet{ymaf05}. The model was produced
with the TYCHO stellar evolution code \citep{ya05}. The model is
non-rotating and includes hydrodynamic mixing processes
\citep{ya05,ymaf05}. Mass loss uses the prescriptions of \citet{kud89}
for OB mass loss and \citet{bl95} for red giant/supergiant mass
loss. The final stellar mass is 14.4 \sol. Figure~\ref{fig1} shows the
mean atomic weight (\={A}, top) and density (bottom) versus enclosed
mass. A 177 element network terminating at $^{74}$Ge is used
throughout the evolution. The network uses the NOSMOKER rates from
\citet{rt01}, weak rates from \citet{lan}, and screening from
\citet{gre73}. Neutrino cooling from plasma processes and the Urca
process is included.

\clearpage
\clearpage

\subsection{Explosion Calculations}

To model collapse and explosion, we use a 1-dimensional Lagrangian
code developed by Herant et al. (1994).  This code includes 3-flavor
neutrino transport using a flux-limited diffusion calculation and a
coupled set of equations of state to model the wide range of densities
in the collapse phase \citep[see][for details]{her94,fryer99a}. It
includes a 14-element nuclear network \citep{bth89} to follow the
energy generation.  This code was used to follow the collapse of the
star through bounce.  For our neutrino driven explosions, we enhanced
the neutrino heating by artificially turning up the neutrino energy.
For these explosions, we found that either we got a very strong
explosion ($10^{52}$\,erg) or no explosion at all.  This was because
if the explosion did not occur quickly, we were unable to drive a
supernova explosion due to the quick decay in the electron neutrino
and anti-neutrino fluxes.  But such a result (either strong or no
explosion) can be overcome if we did not increase the neutrino energy
throughout the star (including the proto-neutron star core) or if we
were adding this heating in a multi-dimensional calculation.  In a
calculation that modifies the neutrino transport by assuming
convection can redistribute the energy in the neutron star and
separately modifies the neutrino transport above the neutron star, one
can easily get a range of explosion energies by increasing the
neutrino energy (Fr\"ohlich et al. 2006).

To get a range of explosion energies, we opted to remove the neutron 
star and drive an explosion by one of two ways: injecting energy 
in the innermost 15 zones (roughly 0.035 \sol) or by driving a piston 
at the innermost zone (outer edge of the proto-neutron star).  The 
duration and magnitude (piston velocity or energy injection) of these 
artificial explosions were altered to produce the different explosion 
energies.  During energy injection, the proto-neutron star is modeled 
as a hard surface.  We do not include the neutrino flux from the 
proto-neutron star, but the energy injected by this neutrino flux is 
minimal compared to our artificial energy injection.  Shortly after 
the end of the energy injection, we turn the hard neutron star surface 
to an absorbing boundary layer, mimicking the accretion of infalling 
matter due to neutrino cooling onto the proto-neutron star.  In this 
manner, we can model the explosion out to late times, even if there 
is considerable fallback.

In this paper, we have focused on the energy-injection model for
explosions.  We have produced 3 sets of explosions (each with a range
of energy) using the same base model where the collapse, bounce, and
fail of the bounce shock is followed with no energy injection (first
380ms after start of collapse).  The first set of models assumes a
fast explosion where the energy is injected in just 20\,ms, producing
an explosion just 400ms after collapse (roughly 150\,ms after bounce).
By varying the energy injected, we produce a range of explosion
energies.  We then modeled two additional sets of explosions where we
injected energy into those inner 15 zones for longer times: 200\,ms
(roughly 330\,ms after bounce) and 700\,ms (roughly 830\,ms after
bounce).  Of course, to produce the same final explosion energy, the
rate of energy input for longer injection times is much lower than
that of our fast explosions.  

The 1D explosions are summarized in Table~\ref{tab1}.  All of our
models assume the same initial mass of the compact remnant: $\sim 1.75
M_\odot$, corresponding to a gravitational mass of the neutron star
roughly equal to $1.6 M_\odot$.  But unless the explosion is extremely
energetic, material falling back onto the proto-neutron star always
produces a much more massive compact remnant, generally producing a
black hole.  We expect such massive progenitors to produce black
holes.  Although it is likely that all stars above 12 \sol have some
fallback \citep{fryer99b}, above 20 \sol, we expect the fallback to be
considerable.  When an explosion is launched, the inner material all
has energy (mostly internal) in excess of the escape energy.  However,
this material deposits much of its energy into the layers above it.
Fallback occurs when this material loses so much energy that it
eventually becomes bound again and {\it falls back} back onto the
compact remnant (see \citet{fk01} for details).

\clearpage
\begin{figure}
\figurenum{2}
\plotone{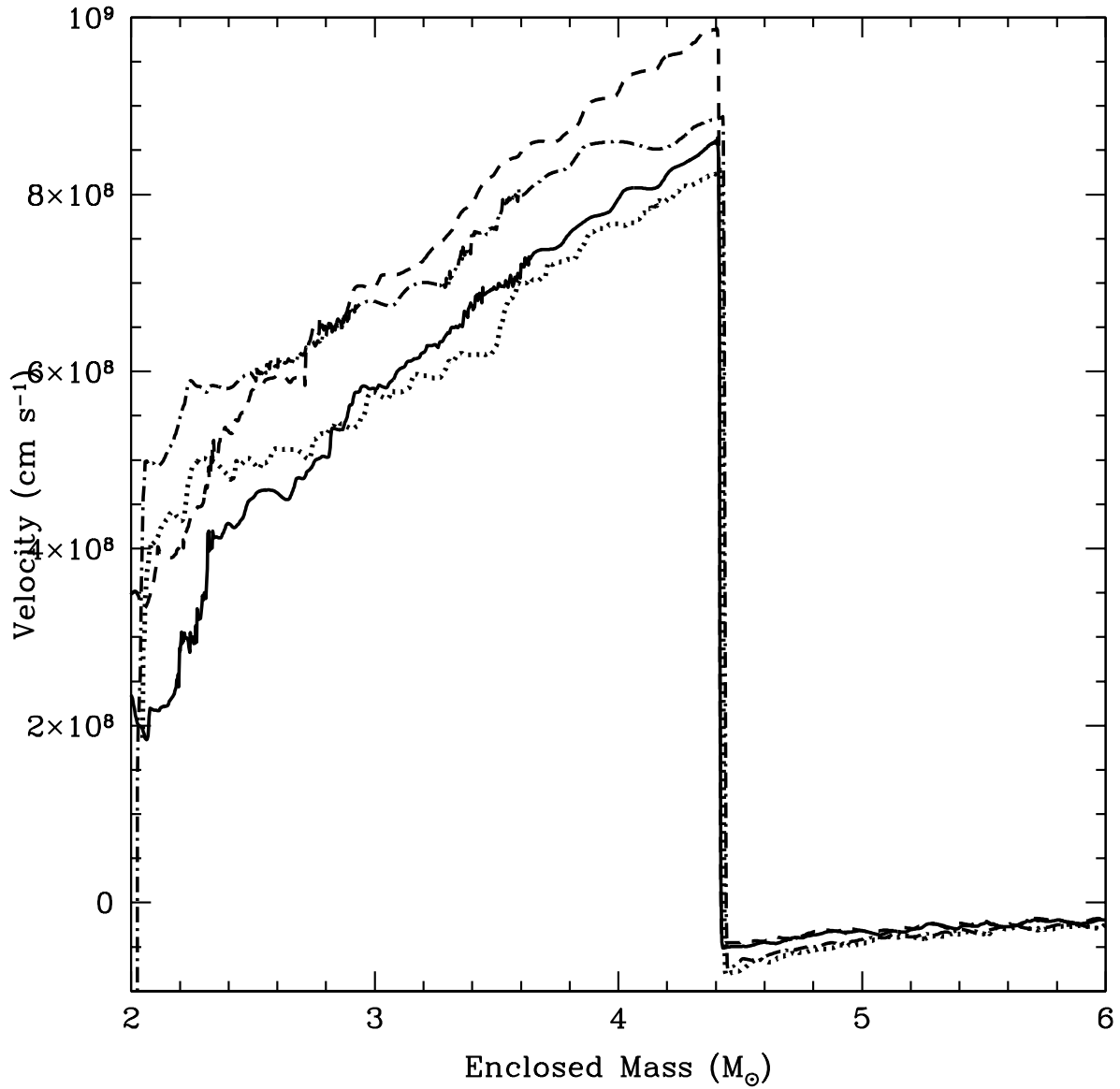}
\caption{Velocity of the ejecta as a function of enclosed mass for two
200\,ms explosions: 23e-0.2-0.8 (solid), 23e-0.2-1.5 (dashed) and two
700\,ms explosions: 23e-0.7-0.8 (dotted), 23e-0.7-1.5 (long-dashed).  
The results are compared when all the shocks are at the same spot.  
This occurs at different times for the different simulations.  
Although the more energetic explosions tend to have higher velocities, 
the duration that the energy is injected also changes the velocity 
profile.
\label{fig1p} }
\end{figure}
\clearpage

Our different explosion methods lead to different amounts of fallback
that, while not physical, allow us probe a range of effects that may
occur in nature.  First off, the piston model moves the material
outward in radius, reducing the potential well that this matter must
climb out to be ejected.  In general, this smaller potential well
leads to less fallback for a given energy than thermal bomb
models. \citet{th97,th02} and Timmes (private communication) note that
there are differences in yields from piston and thermal bomb
explosions. Depending on how one works the piston engine, one could
prevent any fallback at all, although many piston engines do allow for
fallback \citep{fryer99b}. \citet{abt91} find that pistons and thermal
bombs can produce differences in yields on the order of tens of
percent, despite imposing a mass cut to get the desired Fe peak
yields. Comparing the different simulations using a thermal bomb, we
see that the simulations with a longer energy injection also produce
less massive remnants.  Because we continue to inject energy into the
inner cells for longer timescales, these zones move further out of the
potential well and are less likely to fall back.

\clearpage
\begin{figure}
\figurenum{3}
\plotone{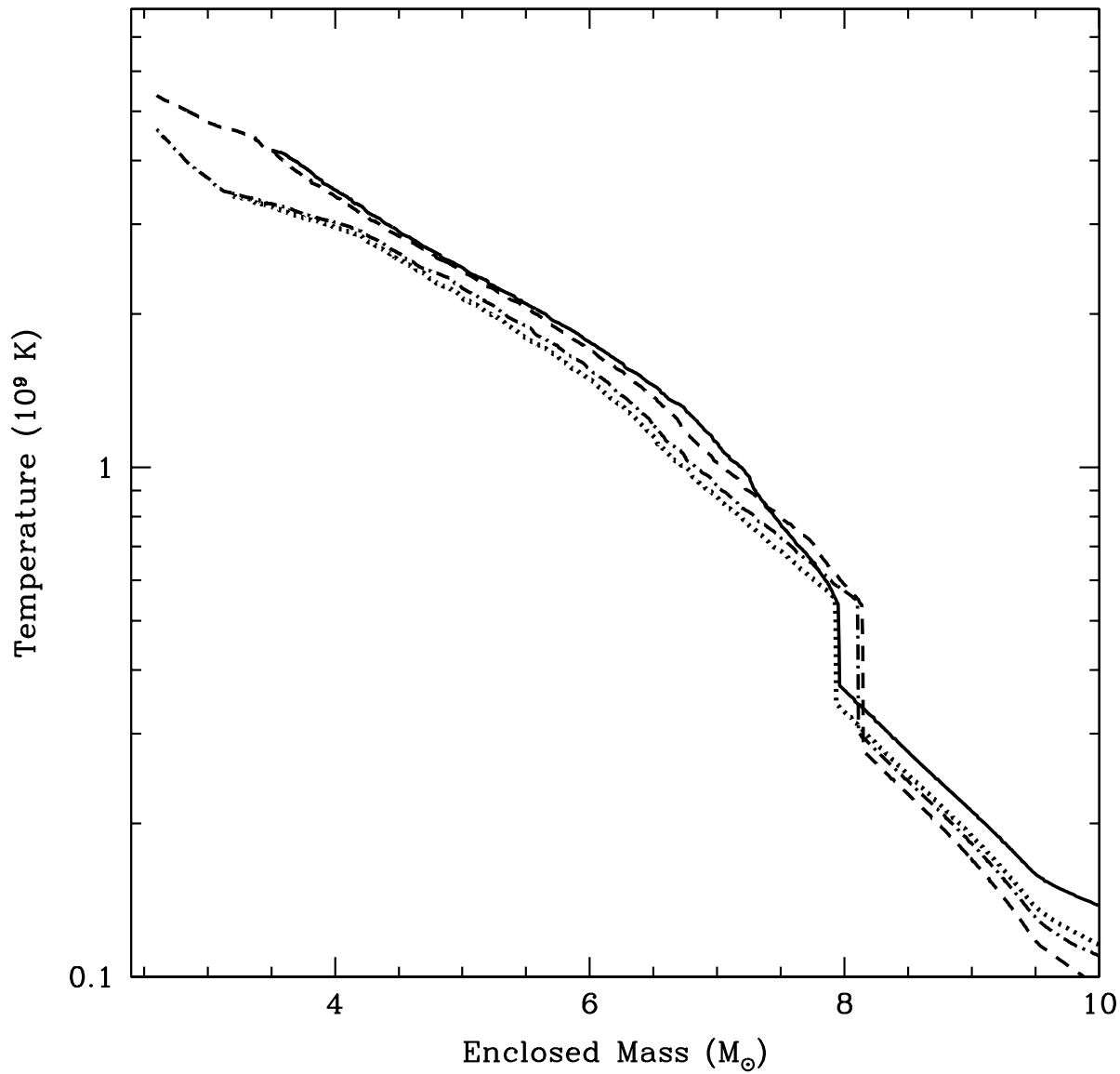}
\caption{Maximum temperature reached by ejecta material as a function
of enclosed mass for two 200\,ms explosions: 23e-0.2-0.8 (solid), 
23e-0.2-1.5 (dashed) and two 700\,ms explosions: 23e-0.7-0.8 (dotted), 
23e-0.7-1.5 (long-dashed).  Note that the weaker shocks in the longer 
duration explosions produced lower peak temperatures.  The difference 
between the maximum temperature based on the explosion energy is 
much less than the difference caused by the duration.
\label{fig1pp} }
\end{figure}
\clearpage

But the remnant mass is not the only difference caused by the duration
of the energy injection.  Figure~\ref{fig1p} shows the velocity
profiles of 4 of our simulations as a function of enclosed mass when
the shock reaches roughly $1.2\times 10^9$\,cm.  Comparing the weak
(0.8 foe\footnote{A foe is a unit made up by Hans Bethe
in his normal sense of humor to denote ten to the {\bf f}ifty {\bf
o}ne {\bf e}rgs ($10^{51}$\,erg).  The foe has also been termed a
``Bethe''.}) explosions (solid, dotted lines) to their normal (1.5 foe)
explosion counterparts, we see the obvious trend: the more energetic
explosion has faster velocities.  But note that the shorter delays in
the 200\,ms explosions produces faster velocities than their 700\,ms
counterparts.  The quicker explosions deposit the energy faster,
leading to a faster initial explosion.  How will this effect the 
nuclear yield?  The shock velocity determines the temperature 
of the gas when it is shocked, so the faster velocities mean 
higher peak temperatures.  The peak temperatures for these models 
is shown in figure~\ref{fig1p}.  Not surprisingly, the quicker 
explosions produce higher peak temperatures.

It should be noted that neither the thermal bomb or piston models are
physically motivated in the sense of representing an actual supernova
mechanism. Both are arbitrary methods of inserting enough energy into
the star to disrupt it.  Although we are far from solving the
supernova mechanism, it is believed that stars in the 18-23 \sol range
lie at the boundary where the neutrino-driven supernova mechanism can
successfully drive strong supernova explosions \citep{fryer99a}.
Following the intuition of the collapse community, one would assume
that this particular star would have weak or no explosions.  This
would severely restrict the range of allowable energies for this star.
However, stars that fail to produce strong supernovae have another way
to explode, the collapsar black-hole accretion disk engine
\citep{woo93}.  This engine can drive extremely energetic
explosions. On the observational side, \citet{mario03} derive a wide
range of energies for Type IIp supernovae. \citet{ken03} find two
branches of supernovae at progenitor masses relevant to this
study. The two populations suggest that two mechanisms might be at
work, one that produces strong explosions and one weak.  Without an
understanding of the explosion mechanism we are free to choose final
kinetic energies for the explosion, and time histories for the thermal
energy deposition or piston.

\subsection{Nucleosynthesis Post-Processing}

The network in the explosion code terminates at $^{56}$Ni and cannot follow
neutron excess so to accurately calculate the yields from these
models, so we turn to a post-process step.  Nucleosynthesis
post-processing was performed with the Burn code \citep{fyh06}, using
a 524 element network terminating at $^{99}$Tc. The network uses the
current NOSMOKER rates described in \citet{rt01}, weak rates from
\citet{lan}, and screening from \citet{gre73}.  Reverse rates are
calculated from detailed balance and allow a smooth transition to a
nuclear statistical equilibrium (NSE) solver at $T > 10^{10}$K. For
this work Burn chooses an appropriate timestep based on the rate of
change of abundances and performs a log-linear interpolation in the
thermodynamic trajectory of each zone in the explosion
calculation. The code also has available modes for analytic adiabatic
trajectories, arbitrary density trajectories coupled with the equation
of state solver in TYCHO \citep{ya05}, Big Bang conditions, and
hydrostatic (stellar) burning. Neutrino cooling from plasma processes
and the Urca process is calculated. The initial abundances are those
of the 177 nuclei in the initial stellar model, and the network
machinery is identical to that in TYCHO.

\section{Yields}

\begin{figure}
\figurenum{4}
\plotone{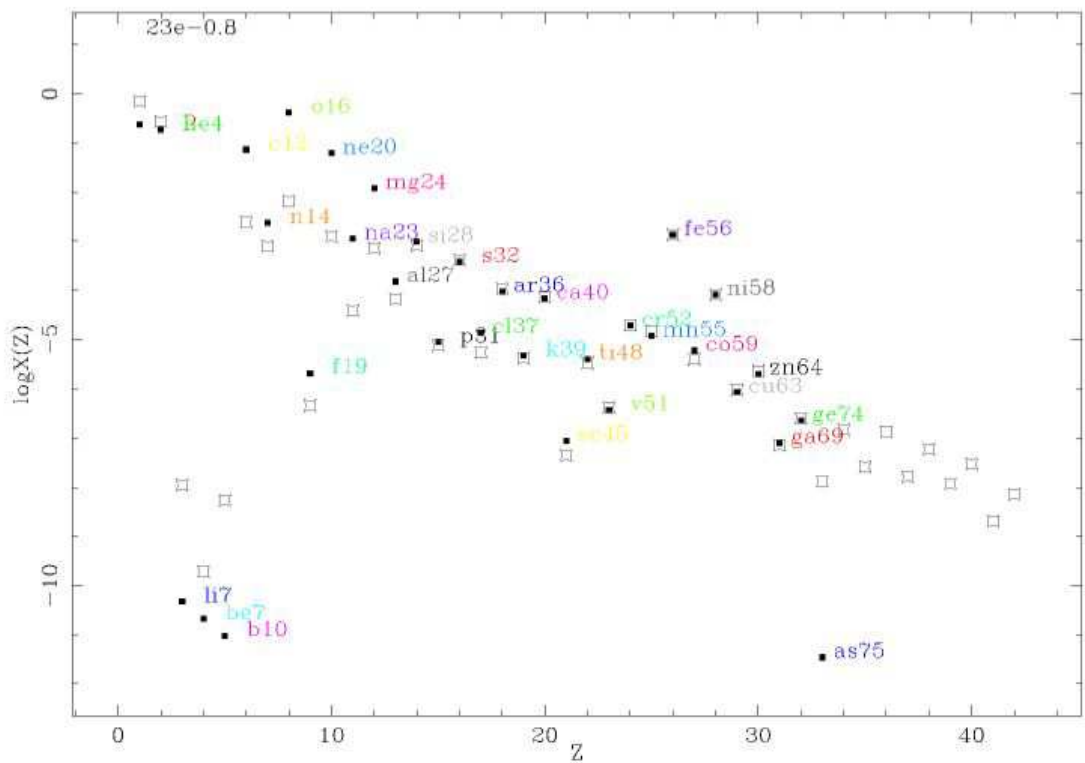}
\caption{Mass fraction of elemental yields $X/X_{\odot}$ versus proton
number Z for the 0.8foe thermal bomb model. Figures~\ref{e1.1}-\ref{e2.0} show
the same for the other thermal bomb explosions) Each element is
labeled with the most abundant isotope of that element. A clear trend
is seen with energy. The 0.8 foe explosion ejects the H, He, and part
of the C/O layers. The heavier elements are represented only by their
initial abundances in the star. As explosion energy increases the
yields of C/O increase, followed by the intermediate mass
elements. Only at 1.5 foe does ejecta reach Si burning temperatures
during the explosion, and produces mostly $^{54}$Fe, characteristic of
high entropy QSE nucleosynthesis. At 2 foe there is ejecta that
reaches NSE conditions and produces $^{56}$Ni as the dominant Fe peak
isotope. \label{e0.8} }
\end{figure}

\begin{figure}
\figurenum{5}
\plotone{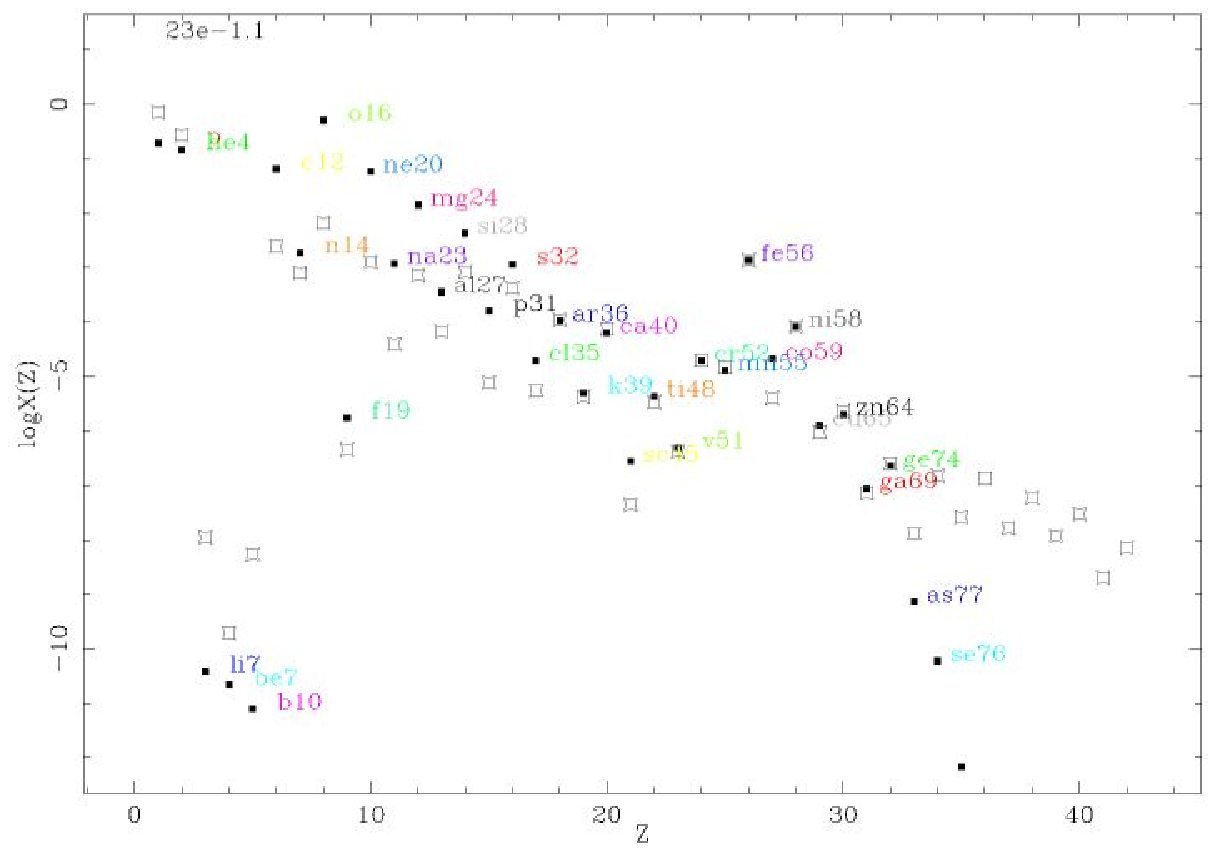}
\caption{Mass fraction of elemental yields $X/X_{\odot}$ versus proton
number Z for the 1.1foe thermal bomb
model. \label{e1.1} }
\end{figure}

\begin{figure}
\figurenum{6}
\plotone{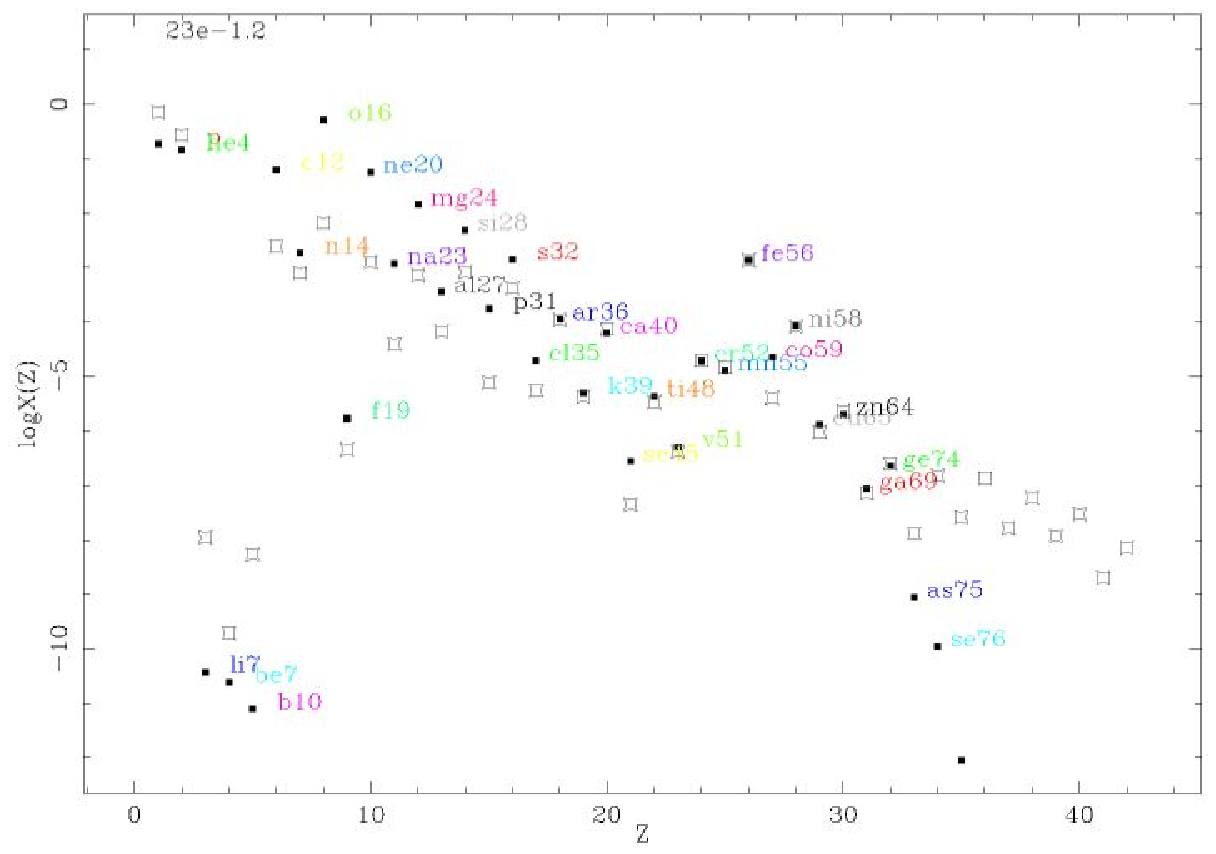}
\caption{Mass fraction of elemental yields $X/X_{\odot}$ versus proton
number Z for the 1.2foe thermal bomb
model. \label{e1.2} }
\end{figure}

\begin{figure}
\figurenum{7}
\plotone{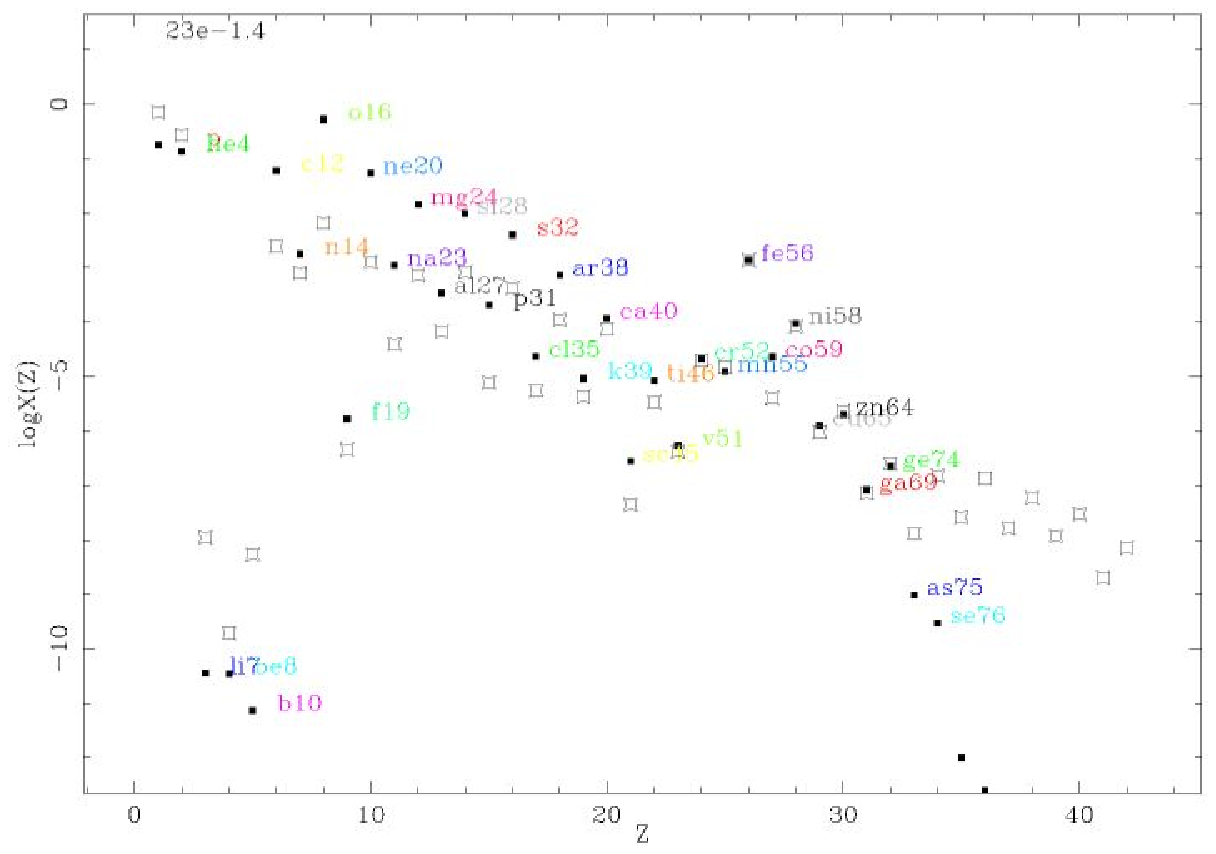}
\caption{Mass fraction of elemental yields $X/X_{\odot}$ versus proton
number Z for the 1.4foe thermal bomb
model. \label{e1.4} }
\end{figure}

\begin{figure}
\figurenum{8}
\plotone{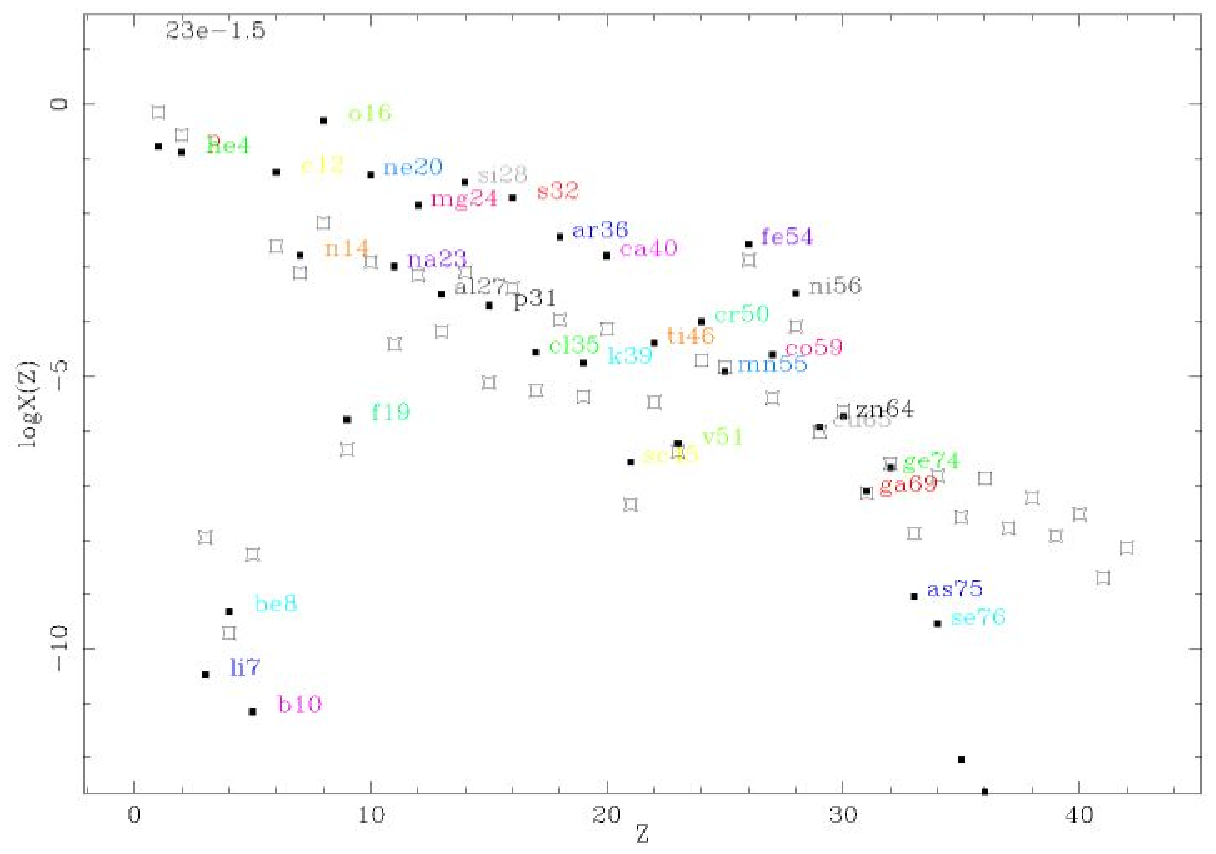}
\caption{Mass fraction of elemental yields $X/X_{\odot}$ versus proton
number Z for the 1.5foe thermal bomb
model. \label{e1.5} }
\end{figure}

\begin{figure}
\figurenum{9}
\plotone{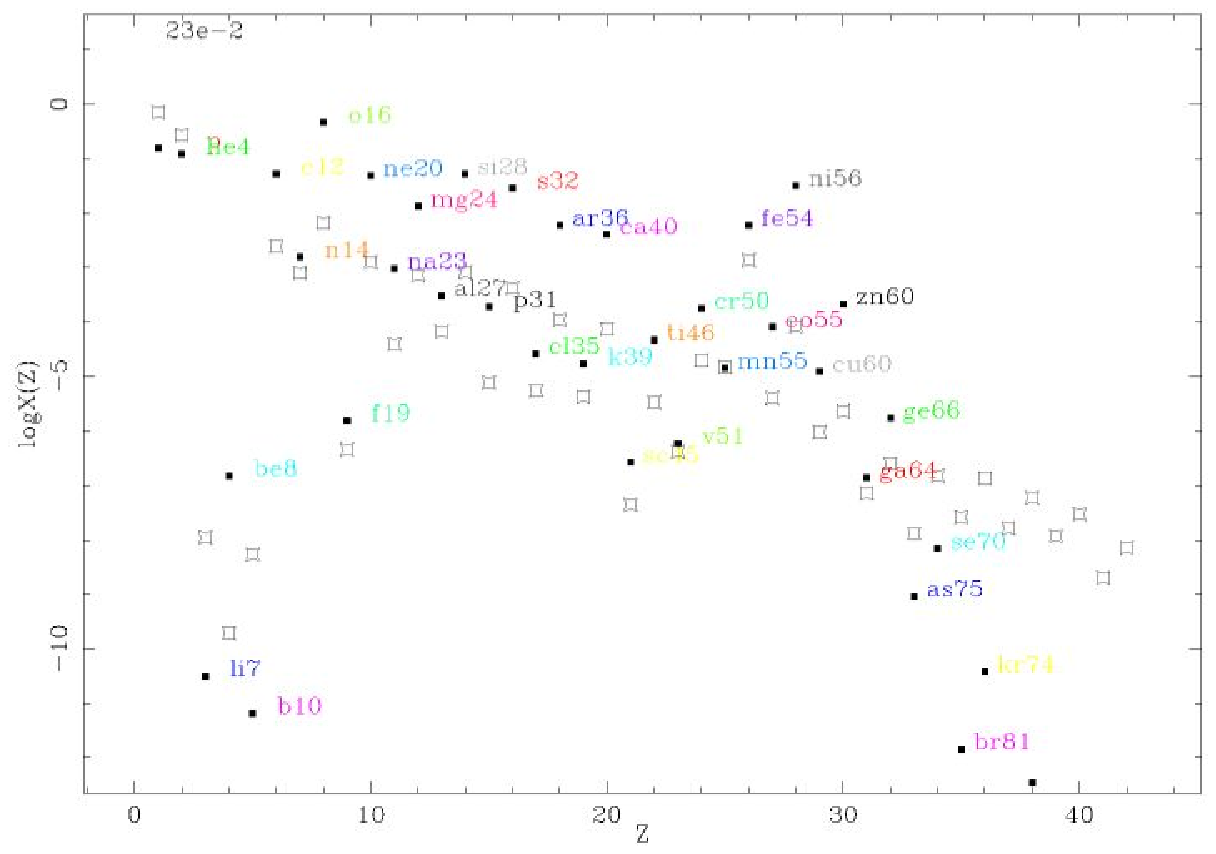}
\caption{Mass fraction of elemental yields $X/X_{\odot}$ versus proton
number Z for the 2.0foe thermal bomb
model. \label{e2.0} }
\end{figure}

\begin{figure}
\figurenum{10}
\plotone{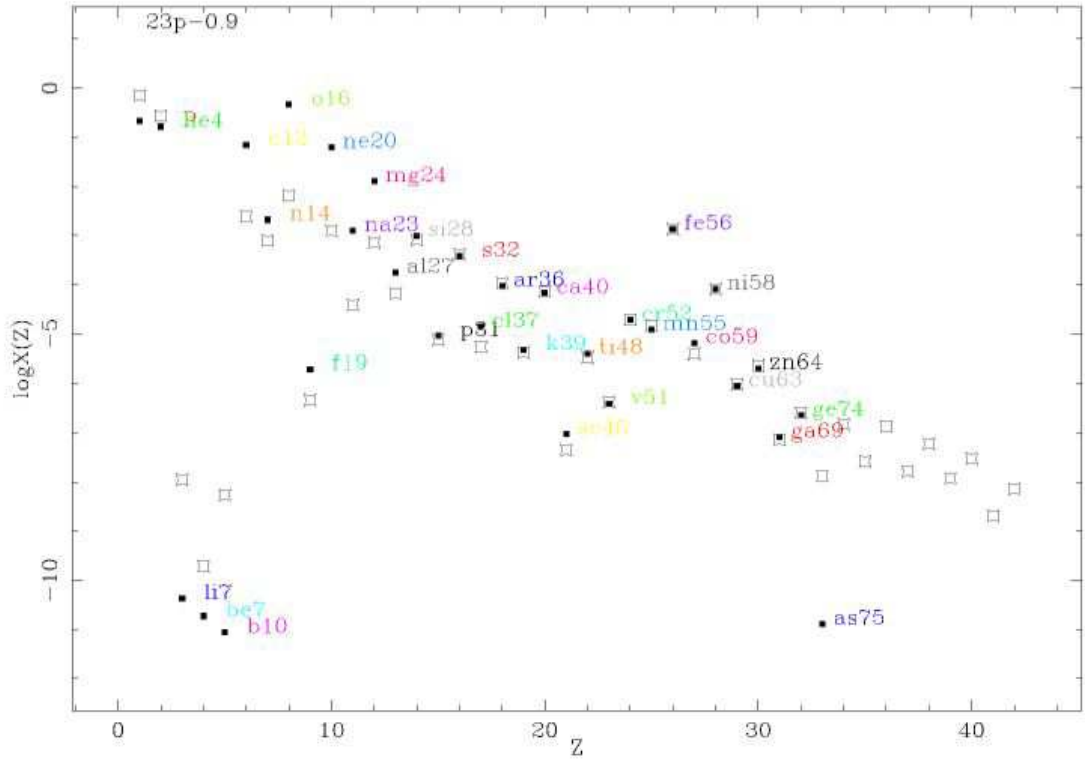}
\caption{Mass fraction of elemental yields $X/X_{\odot}$ versus proton
number Z for the 0.9foe piston model. (Figures~\ref{p1.2} and
\ref{p1.6} show the same for the other piston explosions.) Each
element is labeled with the most abundant isotope of that
element. Trends with energy are similar to the thermal bombs, but
offset to lower energies. \label{p0.9} }
\end{figure}

\begin{figure}
\figurenum{11}
\plotone{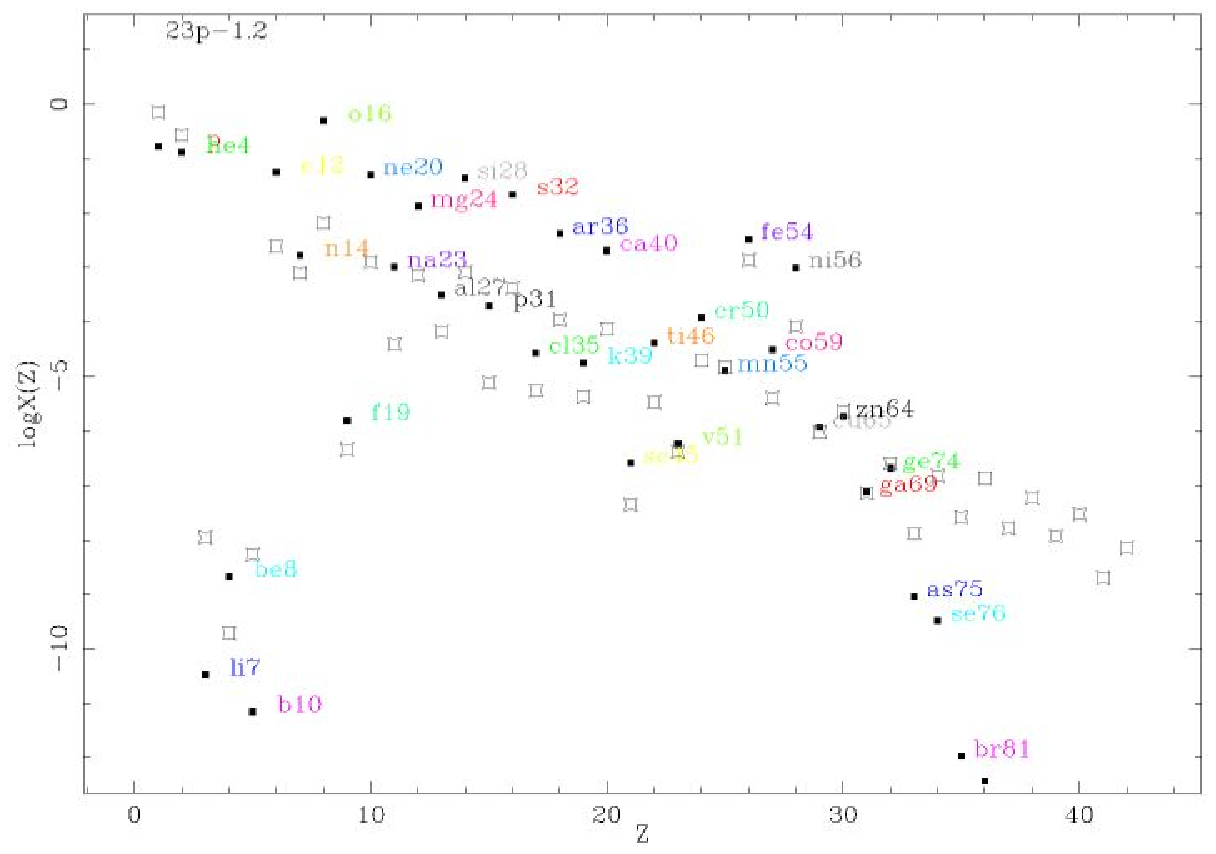}
\caption{Mass fraction of elemental yields $X/X_{\odot}$ versus proton
number Z for the 0.9foe piston model. \label{p1.2} }
\end{figure}

\begin{figure}
\figurenum{12}
\plotone{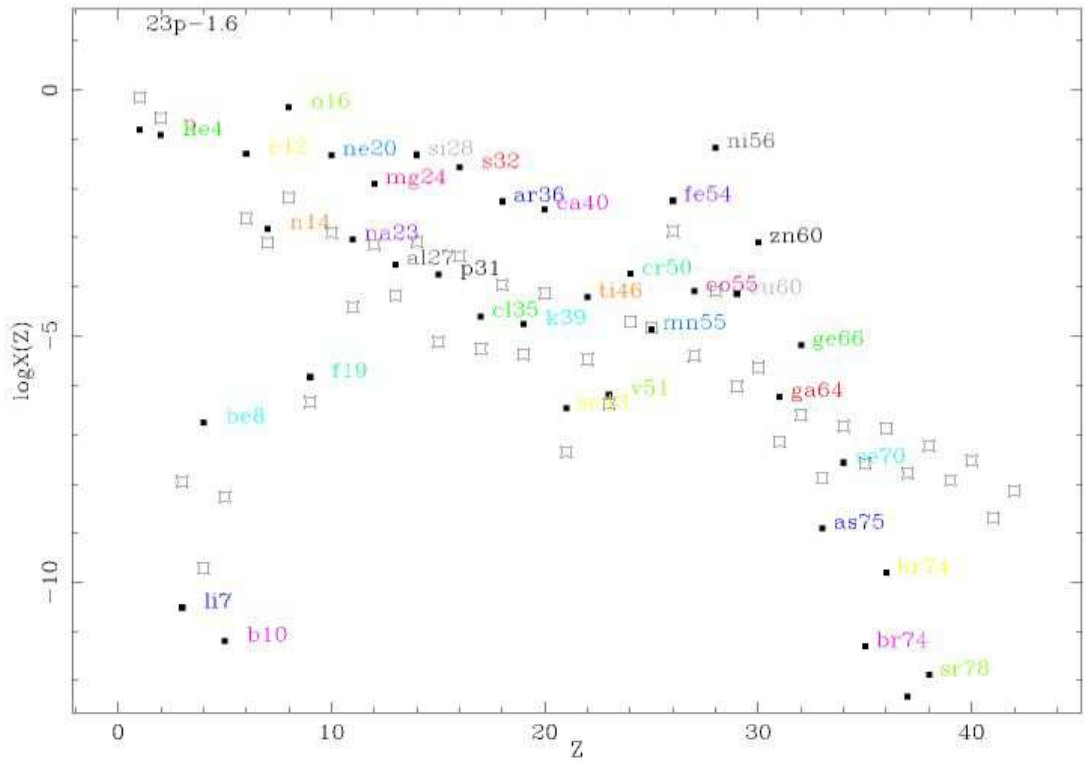}
\caption{Mass fraction of elemental yields $X/X_{\odot}$ versus proton
number Z for the 0.9foe piston model. \label{p1.6} }
\end{figure}

Table~\ref{tab2} and Table~\ref{tab3} give final yields for the
thermal bomb and piston models, respectively. The tables terminate at
$^{74}$Ge. Heavier elements were not tracked during the star's
evolution, so the final yield of those species would not be
complete. The extra extent of the tables does allow us to track the
neutron-rich iron peak correctly, without nuclei we tabulate building
up artificially at a network boundary. In this paper we will
concentrate on the iron peak and intermediate mass elements and defer
any discussion of r and s process to a later
paper. Figures~\ref{e0.8}-\ref{e2.0} show the graphical yields for the
thermal bomb models. Figures~\ref{p0.9}-\ref{p1.6} show yields for
piston models.
\clearpage

\begin{figure}
\figurenum{13}
\plotone{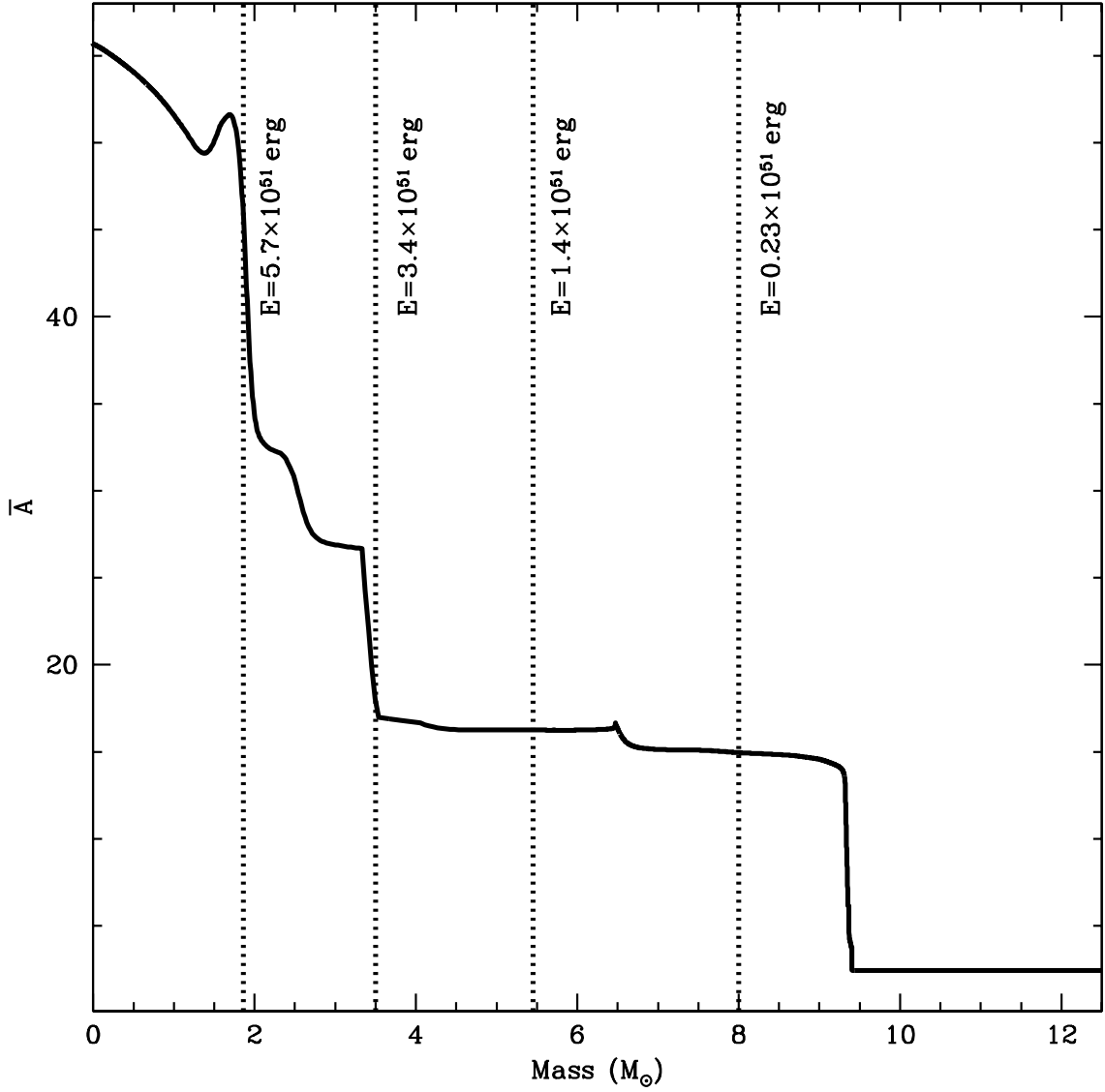}
\caption{Mean atomic mass \={A} vs. enclosed mass in the
progenitor. Dashed vertical lines indicate the material that escapes
for thermal bomb explosions of the indicated energies, ignoring 3D
effects and mixing. \label{escape} }
\end{figure}

\begin{figure}
\figurenum{14}
\plotone{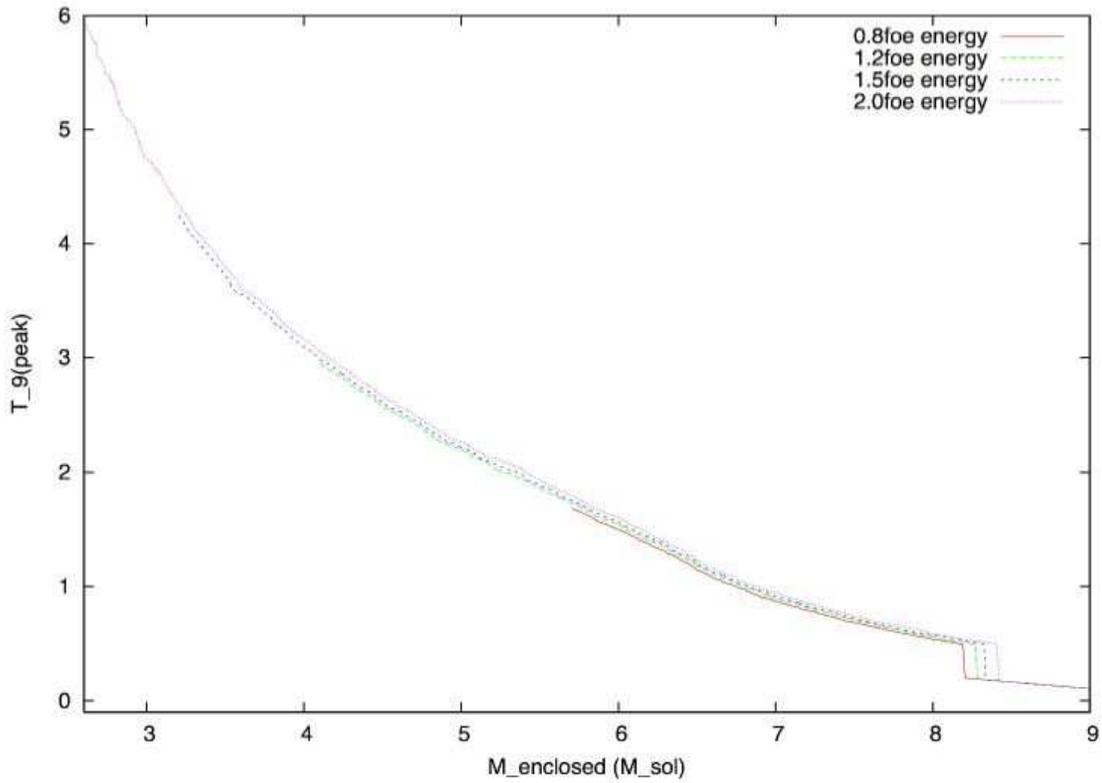}
\caption{Peak $T/10^9$ reached by each mass zone in the ejecta
vs. mass coordinate of the progenitor for the thermal bomb
models. With increasing energy more material escapes and the extra
material is heated to higher temperatures. Only the 2.0 foe explosion
ejects material exposed to NSE conditions. The 1.5 foe explosion
reaches temperatures sufficient for QSE explosive Si
burning. \label{epkt} }
\end{figure}

\clearpage
The thermal bomb models show a clear trend with final explosion
energy. This behavior is simple to interpret. Figure~\ref{escape}
indicates the material which escapes for different thermal bomb
explosion energies on a plot of mean atomic weight \={A} vs enclosed
mass for the progenitor. This ignores of course 3D effects and mixing
during the explosion. Figure~\ref{epkt} shows the peak temperature
reached by each mass zone ejected for the 0.8, 1.2, 1.5, and 2.0 foe
explosions. The explosions with less than 1.4 foe do not reach even
oxygen burning temperatures in the material which escapes as
ejecta. Elements heavier than oxygen are represented only by their
initial abundances in the star. With increasing explosion energy a
larger fraction of the unprocessed material from the original star is
gravitationally unbound, and abundances of intermediate mass elements
begin to come up above solar, progressing towards larger A in order of
synthesis temperature. In the 0.8 foe explosion only about half of the
O-rich mantle escapes. By 1.4 foe Si-rich material makes it out in the
explosion. At 1.5 foe some Si-rich material in the ejecta reaches
quasi-statistical equilibrium temperatures ($>3\times 10^9$K) and is
burned to the Fe peak. In the high entropy QSE regime the material is
dominated by $^{54}$Fe with a smaller component of $^{56}$Ni. At 2.0
foe some of the ejecta reaches full NSE ($T>5\times 10^9$K) and is
dominated by $^{56}$Ni (0.3 \sol).
\clearpage

\begin{figure}
\figurenum{15}
\plotone{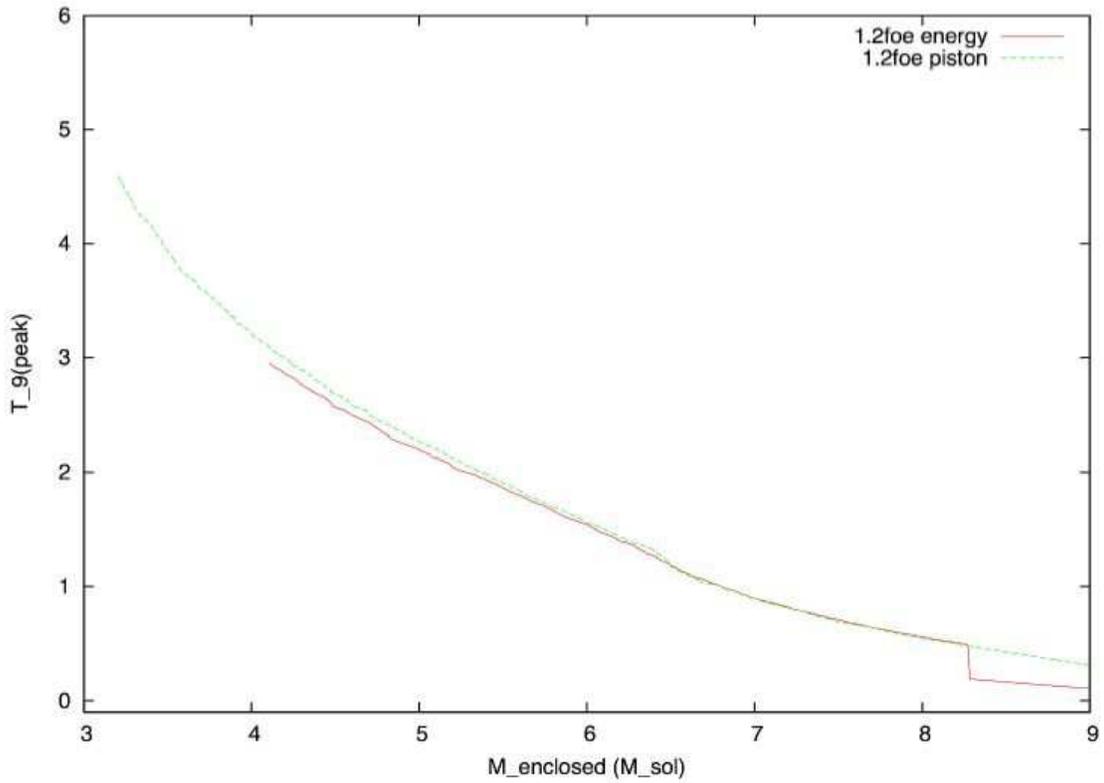}
\caption{Peak $T/10^9$ reached by each mass zone in the ejecta
vs. mass coordinate of the progenitor for the 1.2 foe thermal bomb and
piston models. The difference in yields is due mostly to the larger
amount of material which escapes in the piston explosion. The peak
temperatures for the piston begin to become systematically higher in
the innermost mass zones, but the difference is small for the portion
of the star that escapes in both models. \label{pvepkt} }
\end{figure}

\begin{figure}
\figurenum{16}
\plotone{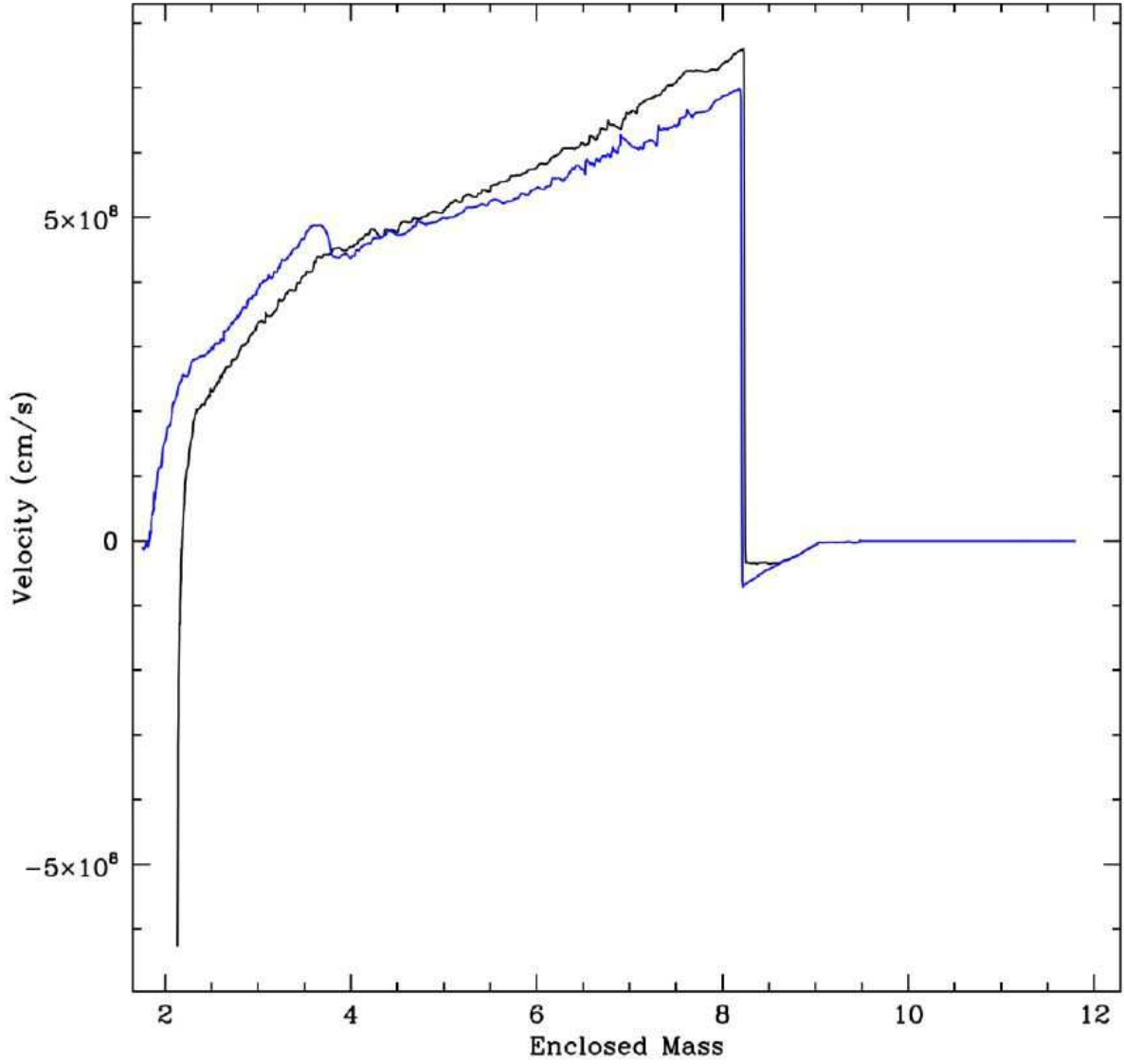}
\caption{Velocity vs. enclosed mass for 1.2 foe piston (blue) and
thermal bomb (black) explosions. The piston model generates large
velocities deeper in the star, resulting in larger ejecta masses and
smaller remnants. \label{vel} }
\end{figure}

\clearpage
This trend is exactly what we already expect with increasing explosion
energy.  At this stellar mass, a range of energies is possible, and we
must consider the entire range of yields.  The more interesting
comparison, however, is between a thermal bomb and a piston explosion
of the same final energy. In terms of yields, the piston explosion
mimics a higher energy thermal bomb explosion because the piston is
more efficient at accelerating material where the explosion is
imposed. Figure~\ref{pvepkt} shows the peak temperature achieved by the
ejecta as a function of initial stellar mass coordinate for 1.2 foe
thermal bomb and piston explosions. Material from deeper in the star
escapes in the piston explosion. Figure~\ref{vel} shows the
velocities versus initial stellar mass coordinate for the 1.2 foe
piston and thermal bomb at the same time during the explosion. High
velocities persist deeper in the star for the piston since it begins
as a mass motion rather than thermal energy that must be transformed
into kinetic energy as in the thermal bomb. 

\begin{figure}
\figurenum{17}
\plotone{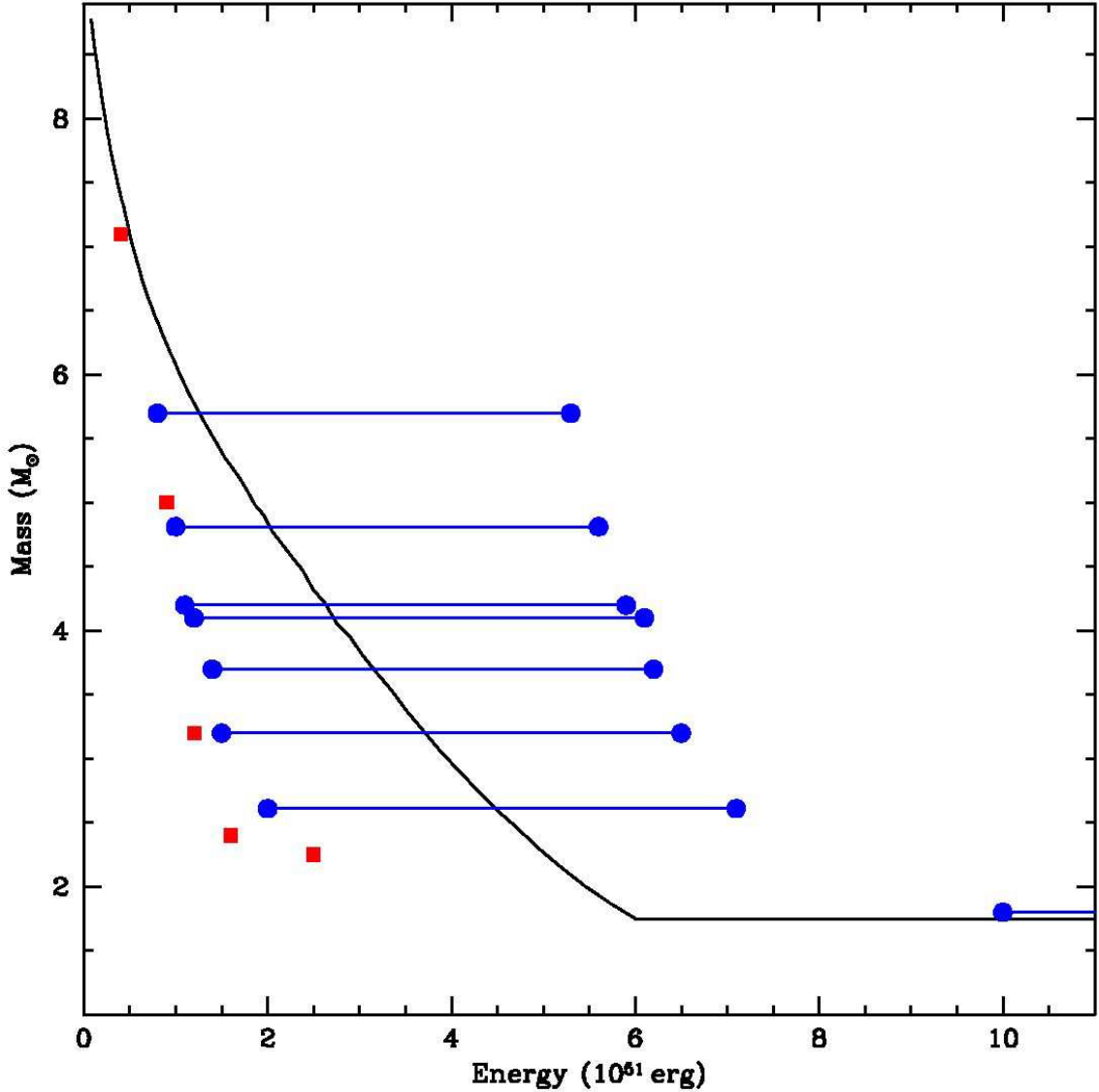}
\caption{Remnant mass vs. energy for thermal bombs (blue pairs),
pistons (red) and analytic estimate (solid line).  The energies for
the piston simulations and the leftmost of each thermal bomb pair are
given in kinetic energy when the shock has broken out of the star.
The analytic estimate and the rightmost point of each thermal bomb
pair gives the energy used to unbind the star, so these two sets of
energies are not directly comparable. Much of the injected energy in
the thermal bombs ultimately is accreted onto the compact remnant and
does not go into exploding the star.  So we expect this energy to
exceed the energy in our analytic calculation.  Piston explosions
systematically produce the smallest remnants for a given final kinetic
energy, due to the efficient initiation of the shock by imposing mass
motion. Larger ejecta masses result in larger yields of highly
processed material. \label{etot} }
\end{figure}

\citet{fryer06} developed an analytic method to estimate the explosion
energy and remnant mass of a collapsing star based on the critical
accretion rate of the infalling star onto the convective region.  By
varying this critical accretion rate, we can determine the dependence
of the remnant mass on explosion energy.  This curve is shown along
with our 1-dimensional results in Figure~\ref{etot}. The pairs of blue
points represent the total energy injected in the thermal bomb models
(rightmost point) and the final kinetic energy of the explosion (left
point). It is difficult to estimate the total energy injected by the
piston. The analytic estimate predicts lower remnant masses for a
given {\it total} energy than the thermal bombs.  The piston model
predicts the lowest remnant masses for a given explosion energy.

There are several differences between the assumptions in the
\citet{fryer06} analysis and the assumptions in our artificially
driven explosions. First, the energy in the analytic estimate is the
energy at the base of the explosion, estimated from the total energy
stored in the convective region.  This is not the total kinetic energy
of the shock as it breaks out of the star.  Much of this energy goes
into unbinding the star. An alternate way to understand this is that
the kinetic energy quoted in our artificial explosion is just a
fraction of the total explosion energy.  It is the total explosion
energy that is quoted in most core-collapse engine calculations, but
it is the final kinetic energy that is quoted in most estimates based
on observations.  So the primary difference between these results is
the difference between the definitions of explosion energy between
these two models.  It is difficult to estimate the energy injected in
the piston model, but we can calculate the energy injected in our
thermal bomb calculations.  Unfortunately, much of the energy injected
in our artificial explosions ends up falling back onto the compact
remnant, so the energy we deposit is likely to be higher than our
analytic analysis that assumed all of that energy escaped.  We see
that this is true for all of our simulations (Figure~\ref{etot}).

Other differences also exist.  In the thermal bomb and piston
explosions, the engine continues to be powered even after the
explosion is launched.  In both the analytic estimate, and in a
neutrino-driven explosion, the explosion energy is limited to what can
be stored in the convective region before the the launch of the
explosion.  After this launch, neutrinos will not deposit much more
energy into a strong explosion.  The continued driving that occurs in
our artificial explosions leads to less fallback than we would expect
from neutrino-driven explosions.  In this respect, the analytic
estimate is more realistic than our 1-dimensional simulations.
\clearpage

\begin{figure}
\figurenum{18}
\plotone{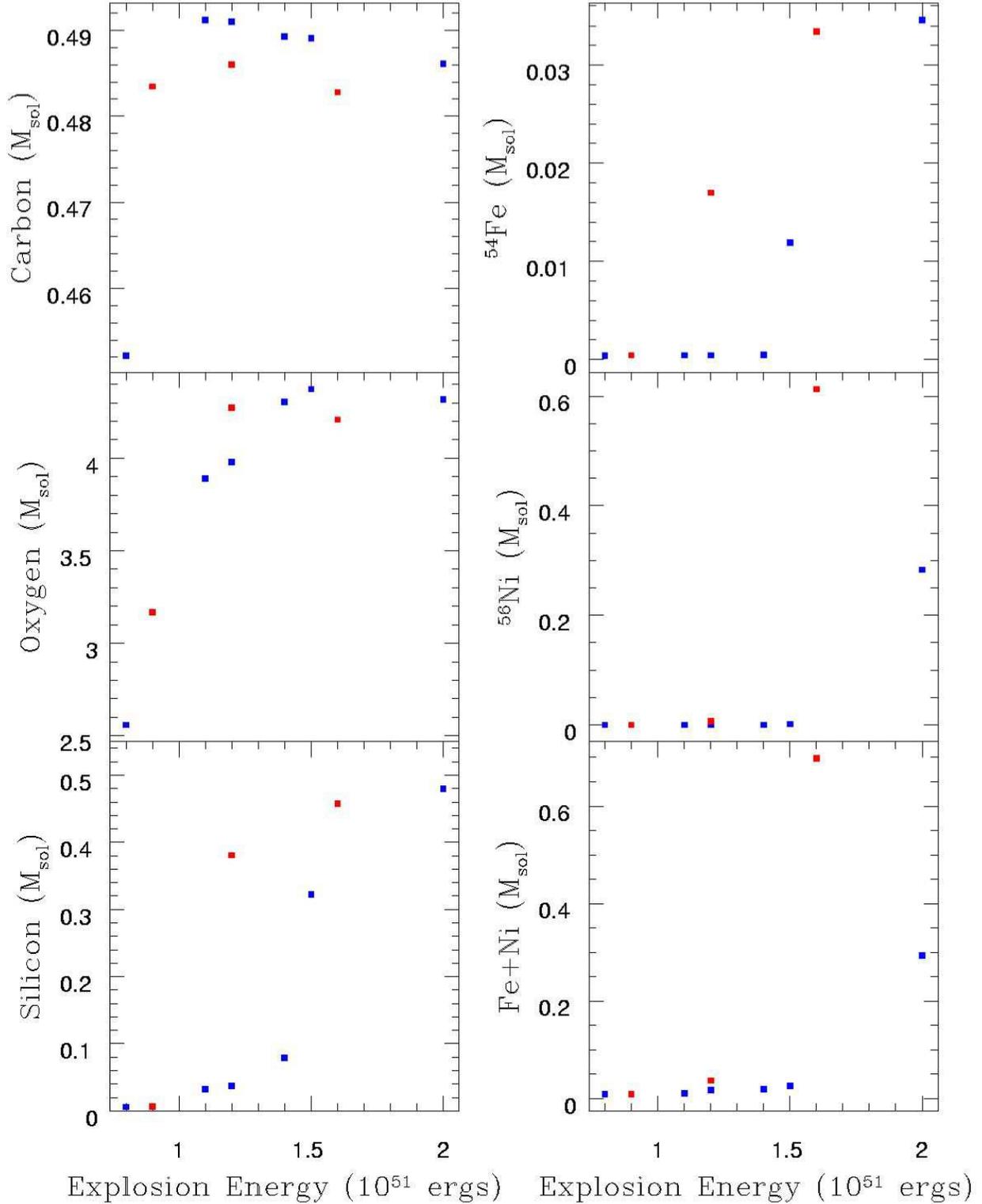}
\caption{Yields (\sol) of carbon, oxygen, silicon, $^{54}$Fe,
$^{56}$Ni, and the total iron plus nickel as a function of explosion
energy, with thermal bombs in blue and pistons in red. \label{pveyield} }
\end{figure}

\clearpage
Over the energy range 0.8 - 1.5 foe in the thermal bombs the yield of
oxygen increases by 70\%, neon by 17\%, silicon by 530\%, and
iron+nickel by 290\%. We can also compare thermal bomb and piston
explosions of the same energy. At 1.2foe the yields in the piston
explosion increase as follows: oxygen 8\%, Ne 0.5\%, Si 1011\%, Fe+Ni
324\%. The $^{56}$Ni yield increases from essentially zero to
6.6$\times 10^{-3}$\sol. Figure~\ref{pveyield} shows the yields (\sol) of
carbon, oxygen, silicon, $^{54}$Fe, $^{56}$Ni, and the total iron plus
nickel as a function of explosion energy, with thermal bombs in blue
and pistons in red. Carbon decreases at higher energies as more ejecta
is heated to C burning temperatures. Oxygen increases as a larger
fraction of the O-rich mantle is ejected, then decreases at energies
that produce O burning temperatures. Silicon has a sharp turn-up where
pre-supernova Si is finally ejected. $^{54}$Fe rises rapidly for
energies that produce Si burning ($T > 3\times 10^9$K) and then drops
at energies high enough for the equilibrium to favor $^{56}$Ni ($T >
5\times 10^9$K). Total Fe+Ni of course increases rapidly at energies
that produce Si burning temperatures. These trends are the same for
both types of explosion, but are shifted to lower energies for the
piston models. The effect is especially dramatic for the Fe peak.

\clearpage

\begin{figure}
\figurenum{19}
\plotone{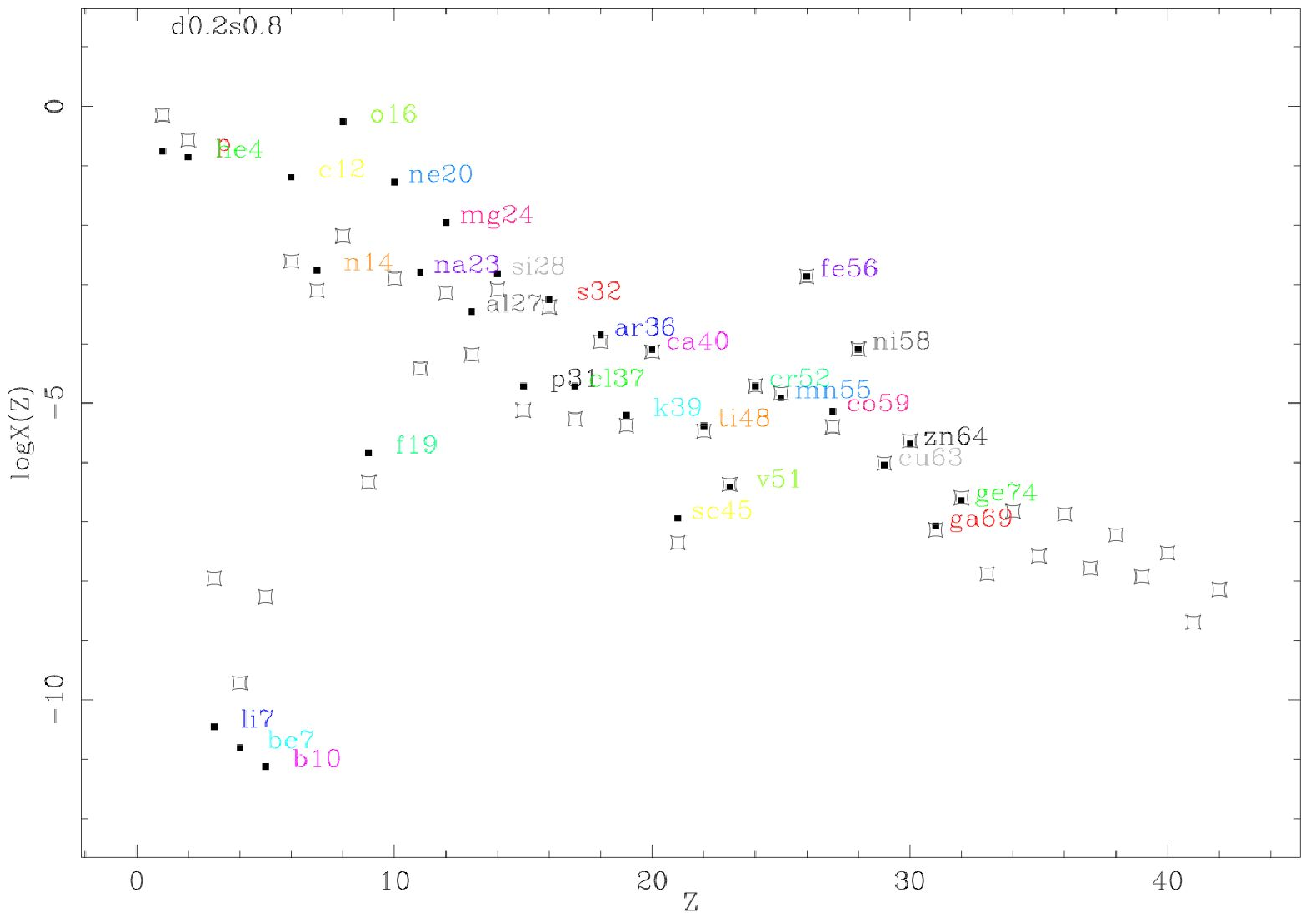}
\caption{Mass fraction of elemental yields $X/X_{\odot}$ versus proton
number Z for the 0.8foe thermal bomb
model with an additional 0.2s delay. \label{d0.2s0.8} }
\end{figure}

\begin{figure}
\figurenum{20}
\plotone{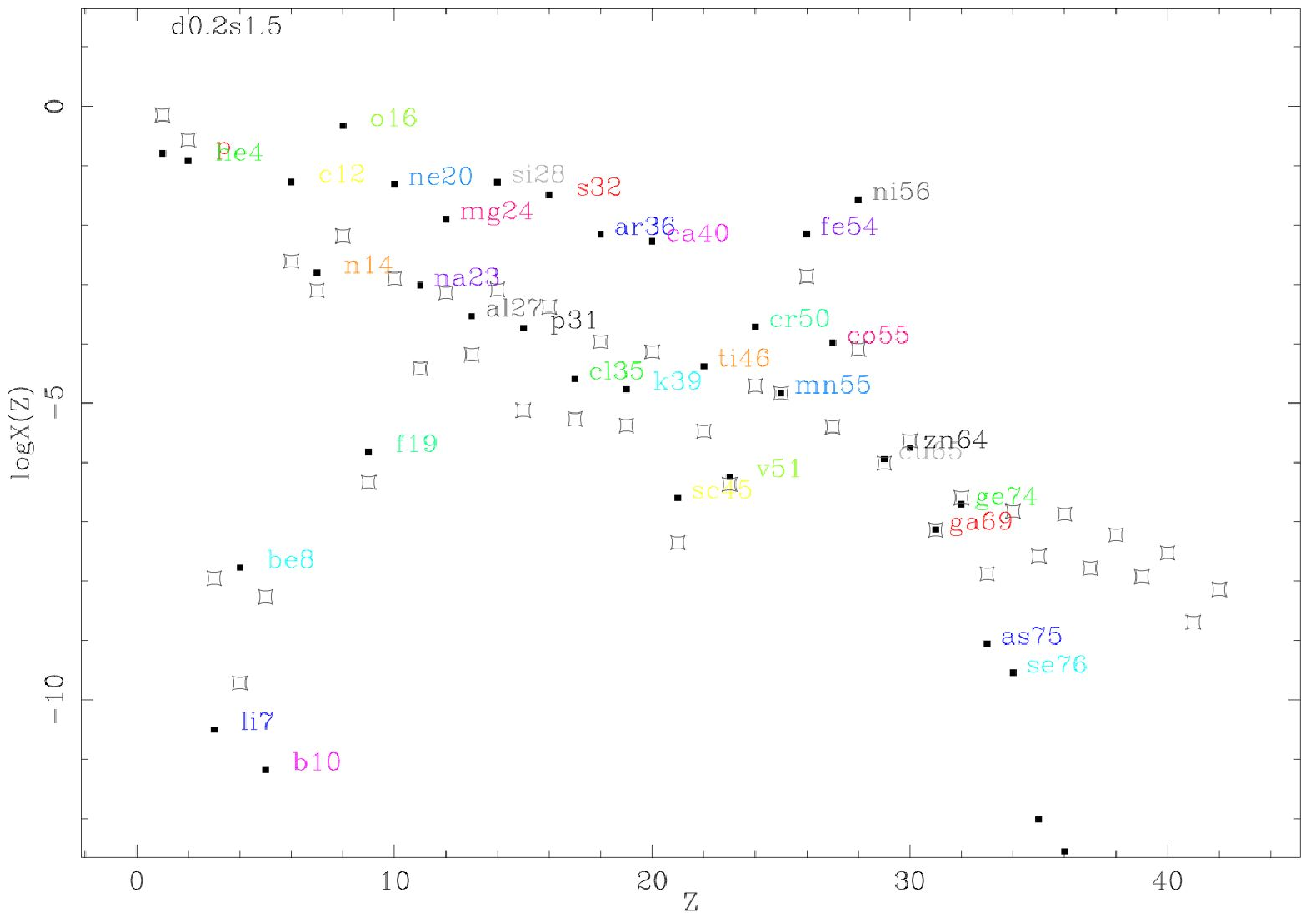}
\caption{Mass fraction of elemental yields $X/X_{\odot}$ versus proton
number Z for the 1.5foe thermal bomb
model with an additional 0.2s delay. \label{d0.2s1.5} }
\end{figure}

\begin{figure}
\figurenum{21}
\plotone{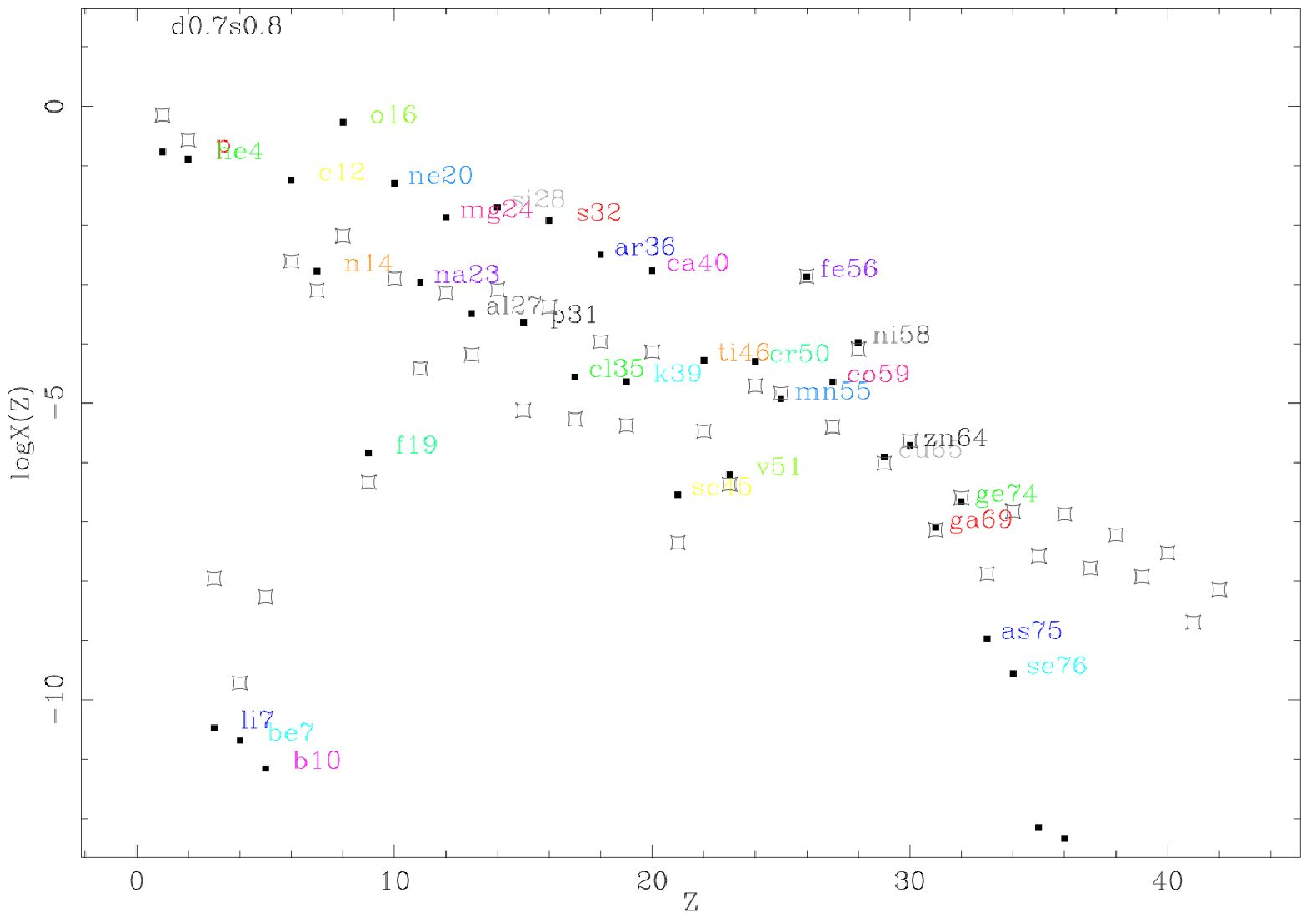}
\caption{Mass fraction of elemental yields $X/X_{\odot}$ versus proton
number Z for the 0.8foe thermal bomb
model with an additional 0.7s delay. \label{d0.7s0.8} }
\end{figure}

\begin{figure}
\figurenum{22}
\plotone{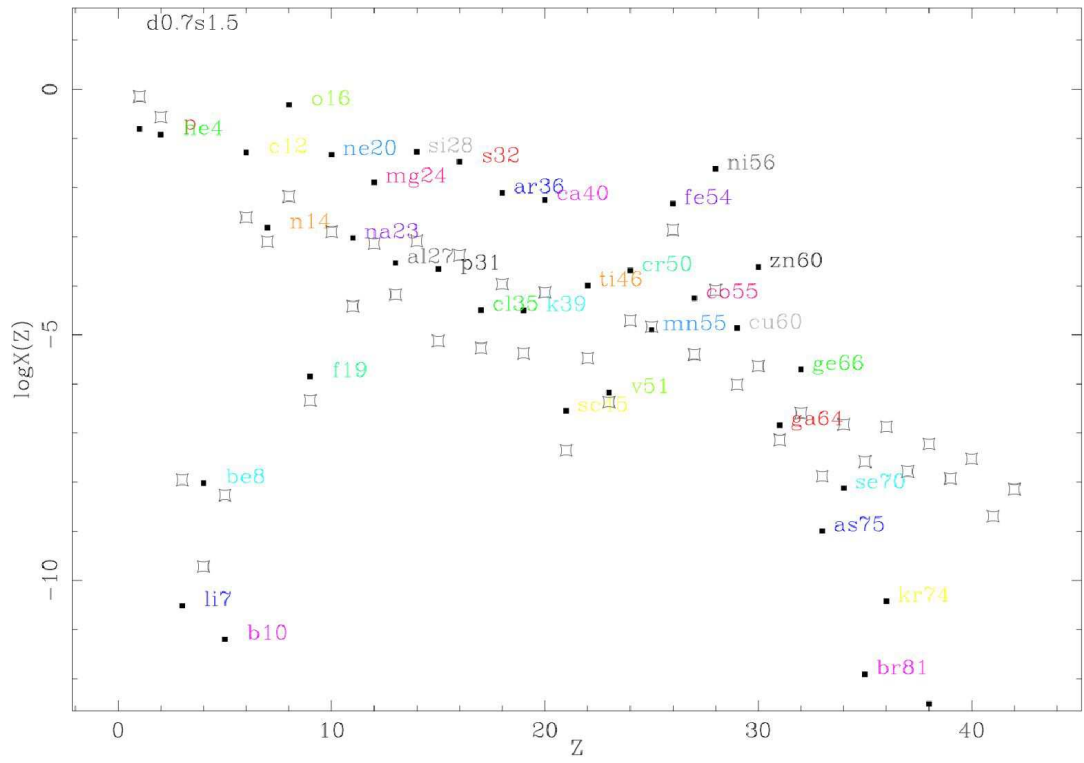}
\caption{Mass fraction of elemental yields $X/X_{\odot}$ versus proton
number Z for the 1.5foe thermal bomb
model with an additional 0.7s delay. \label{d0.7s1.5} }
\end{figure}

We also compare thermal bomb explosions with different explosion
delays, as explained in Section 2.2. For a star this massive it is
reasonable to expect a long delay in light of 3D simulations
\citep{fy06}. Figures~\ref{d0.2s0.8}-\ref{d0.7s1.5} show the bulk
element yields for explosions with additional delays of 0.2 seconds at
0.8 and 1.5 foe and 0.7s at 0.8 and 1.5 foe. These can be compared to
Figures~\ref{e0.8},\ref{e1.5},\ref{p0.9},\ref{p1.6}. The detailed
isotopic yields are presented in Table~\ref{tab4}. We find that for
longer delays the yield of intermediate mass and Fe peak elements
increases because the longer energy input raises more material out of
the potential well. However, this material is at systematically lower
velocities and thus lower shock temperatures (see Section 2.2). As a
result, the yield of Fe peak material processed at NSE temperatures is
lower in the 1.5 foe explosions with longer delays than the 2.0foe
short delay, even though as much or more total material escapes. The
$^{56}$Ni yield increases by two orders of magnitude for both moderate
and long delays. $^{44}$Ti increases by a factor of five with 0.2s
additional delay and by another factor of five with 0.7s as more
material is heated to Si burning QSE and escapes with longer
delays. These results should be taken as indicative of the range of
yields that can be generated with variation of parameters than a real
trend. In 3D a longer delay allows {\it more} fallback, so this may be
the opposite of what nature does.

\clearpage
\section{DISCUSSION}

This research provides a framework for estimating uncertainties in
yields arising from the shortcomings of 1D explosion models. These
errors fall roughly into two regimes. 

In the low explosion energy regime the variation in yields arises
simply from the amount of material processed by hydrostatic burning
which escapes from the star. Ignoring multi-D effects (the importance
of which should not be underestimated), this can be described by a
traditional mass cut. 

In the high energy explosion regime we must deal with explosive
nucleosynthesis as well. In this case the thermodynamic trajectory of
the ejecta is important. The velocity evolution of a piston explosion
and a thermal bomb of the same energy or of a piston and a thermal
bomb with the same mass cut will differ slightly, imposing an
accompanying difference in the thermodynamic evolution. Hence any
material which undergoes explosive nucleosynthesis will have different
abundances. The 1.2foe piston and 1.5foe thermal bomb produce the same
mass compact remnant. If we examine the elemental yields we see that
the effect is small, as the difference in the velocities is small. In
individual isotopes, however, the difference can be larger. $^{56}$Ni
is a factor of three larger in the piston driven explosion because its
production is dominated by the highest temperature material, in which
the largest deviation in thermodynamic histories is seen. This changes
the Fe yield by $\sim$25\%. The evolution would be substantially
different for an explosion with a gain region as in the analytic case,
and the abundances correspondingly disparate. A small change in
energy/compact remnant mass in this regime results in a much larger
difference in the yields than a similar change in the low energy
regime due to the steep temperature dependence of the burning. In
addition, the time delay of the explosion has an effect in this
regime. A longer delay {\it in 1D} results in more material escaping,
but with lower shock temperatures, resulting in a higher intermediate
mass to Fe peak ratio. In 3D, we would expect more fallback, resulting
in a lower overall ejecta mass.

We can choose between two parameters in constructing a grid of
yields. If we choose to parameterize by compact remnant mass, which
makes more sense in terms of yields, we must still take into account
the thermodynamic history of the material which escapes. If we
parameterize by explosion energy, which has some hope of being
partially constrained by observations of explosions, we are left with
a very uncertain remnant mass, amount of ejecta, and thermodynamic
history.

\citet{fk01} examine how supernova explosion energy varies with
mass. They argued that the supernova explosion energy peaks at an
initial progenitor mass of 15M$_\odot$.  At 23M$_\odot$, they
predicted an explosion energy of less than $10^{51}$\,erg. While the
progenitors we have used in this paper are very different from the
progenitors used in \citet{fk01}, we may still use the trends in that
work to guess where the parameter space will open up for the
uncertainties identified here. We find that large explosion energies
are required to expel explosively processed material at 23 \sol
because of the massive core of the star. It is likely that
sufficiently large explosion energies only occur if the collapsar
engine is driving the explosion.  If such stars do have low explosion
energies, then our primary uncertainty will be in the O/Ne/Mg to Si/S
part of the periodic table, where we may expect changes in O of tens
of percent and changes in Si of factors of a few. In this case it is
sufficient to examine a range of total ejecta masses to estimate the
range of yields, remembering that the energy corresponding to a given
remnant mass differs by $\sim$30\% between piston and thermal bomb
models and is even larger for a neutrino gain region. As our
progenitor decreases in mass, explosion energy rises and the amount of
energy required to eject explosively processed material declines. By
analogy to the models in this work, the mass range where our expected
explosion energy crosses into the regime where QSE and NSE processing
take place can have uncertainties in Si by a factor of 10 and in Fe
peak by factors of several. Useful explosion diagnostics like
$^{56}$Ni can vary by orders of magnitude. At higher explosion
energies the main uncertainties lie in the Fe peak, with a smaller
contribution in the amount of Si rich material left unburned. The
situation is rendered more difficult by the fact that this
parameterization relates progenitor mass to explosion energy, which
does not have a one-to-one correspondence with remnant/ejecta mass.

We expect our maximum uncertainties for lower mass, and thus more
common, progenitors. For these stars a two parameter grid in remnant
mass and thermodynamic history must be calculated to estimate the
uncertainties. Without a good understanding of the supernova mechanism
and the remnant to explosion energy relation, these grids and their
corresponding error bars must be large. Perhaps more importantly, the
solutions for a given abundance pattern are not unique, especially if
we measure only elemental abundances. Ignoring variations in
progenitors, initial mass functions, and 3D complications, we can
independently adjust the explosion energy distribution, the
energy/remnant mass relation, and the thermodynamic history of an
explosion to find a given result. To some extent we can constrain our
parameter space by attempting to match the solar abundance pattern
(which is itself in dispute! \citep{ap05}), but this gives us very
little leverage in making predictions since it introduces further
complications of star formation histories and the non-linearity of
enriching intermediate generations of model stars with our own
uncertain yields.

It is worth discussing these results in the context of previous
studies. We will begin with \citet{ww95}, which provides the current
standard for yields and the template from which later compilations of
yields are derived. Globally, we notice that varying the explosion
energy or the treatment of the explosion can have very large effects
not only on individual, low abundance diagnostic isotopes, but also on
the bulk yields of common elements.  \citet{ww95} examine different
effective gravitational potentials in the piston (broadly equivalent
to changing the final explosion kinetic energy while keeping the same
energy input) for stars $>$ 30 \sol. Below this mass limit they
explore some small changes in the method of implementing the piston,
which do have effects on important low abundance isotopes. However,
they do not change the bulk energetics, assuming a final explosion
kinetic energy of $\sim 1.2\times 10^{51}$ erg.  \citet{ww95} do point
out that the energies are uncertain, but do not present any resulting
uncertainties in the yields. If anything, the assumption of constant
energy looks worse now. Our still unsatisfactory understanding of the
mechanism at least assures us that the explosion energies for stars in
the common 10-20 \sol range will vary a great deal. Uncertainties of
factors of more than ten in silicon and the iron peak can occur for
stars of masses common enough to be interesting to population yields,
given what we must currently accept as a reasonable range of explosion
energies.

More specifically, we can compare our 1.2 foe piston to the
\citet{ww95} S22A model. Our yield of carbon is almost exactly twice
that of S22A. Oxygen is increased by $\sim 80$\%, Si by 10\%. The S22A
Fe peak is nearly a factor of seven higher than our 23p-1.2, due
almost entirely to their much larger yield of $^{56}$Ni. Almost all of
these changes can be traced to differences in the progenitor
model. Our progenitor includes mass loss and, most importantly, a much
more realistic treatment of hydrodynamics and mixing, which results in
much larger core sizes for a given mass. This contributes to the
$^{56}$Ni discrepancy as well, but as \citet{ww95} point out, this is also
sensitive to the details of the piston.

\citet*{abt91} examine in detail two different methods of launching a
1D explosion with thermal bombs and pistons, and so provide an
interesting comparison to this work. Their results at first glance
appear to diverge significantly from ours, but when the differing
assumptions of the two calculations are taken into account we shall
see that the results of both studies are quite consistent. The
following discussion highlights some of the vagaries of using 1D
explosion models.

\citet*{abt91} find that the temperature profiles for thermal bombs and
pistons are nearly identical in most of the stellar mantle, beginning
to diverge increasingly, starting at several percent when $T$ exceeds
$2\times 10^9$K (see their Figure 7). The mass represented by very
large temperature differences is small. Since they define the ejected
material by imposing a uniform mass cut, the bulk yields for a given
final kinetic energy are similar.

The principal difference in our results versus those of \citet*{abt91}
results from our treatment of the hydrodynamic evolution. We do not
impose a mass cut; rather we let material without sufficient kinetic
energy to become unbound fall back onto the compact object. If we
return to Figure~\ref{pvepkt}, which compares the peak temperatures in
thermal bomb and piston models with a final kinetic energy of 1.2foe,
we see that the temperatures in the ejected material begin to diverge
at about $T > 2\times 10^9$K, but the difference remains small for the
material ejected, just as in the \citet*{abt91} study. The two models
have very different yields, however, because much more material falls
back in the thermal bomb case. Indeed, when we compare the 1.2foe
piston and 1.5foe thermal bomb or 1.6foe piston and 2.0foe thermal
bomb, which have similar compact remnant masses, we find that the
yields are reasonably similar, with the exception of NSE species such
as $^{56}$Ni, which we would expect to show the effects of the initial
difference in partition of kinetic and thermal energy between the
thermal bomb and piston. Given that we are free to specify the
location of the piston, it's acceleration, velocity, and duration, we
could probably produce more similar kinetic energies for a given
compact remnant mass, but this {\it ad hoc} fine tuning obscures the
amount of variation produced by taking an independent ``best guess''
in setting up the two explosions to reproduce the observable
constraint of ejecta kinetic energy.

The other notable difference is the mass coordinate at which
corresponding temperatures occur in each study. \citet*{abt91} use a
20\sol stellar model from \citet{ken90}. Our progenitor is a 23 \sol
model with hydrodynamic that accounts for hydrodynamic mixing
processes that produce much larger nuclear processed cores and burning
shells. The much more massive \={A}$ \ge 16$ mantle in our model
requires much more energy input to produce a successful explosion, so
high temperatures extend to higher mass coordinates, as do the Fe core
and O and Si shells. This is exactly as we would expect. Once adjusted
for the differences in progenitors, the agreement of our results is
surprisingly good.

More recent studies corroborate these results. \citet{un05} examine
nucleosynthesis in population III stars with varying amounts of
fallback and mixing. While it is obviously impossible to compare our
results directly, since the progenitors are of entirely different
mass, metallicity, and input physics, their methodology has some
bearing on this study. Their figure 14 is particularly interesting,
showing the allowed range of fallback masses required to derive a
Mg/Fe ratio consistent with extremely metal-poor stars. They choose an
initial mass cut of 2.44\sol, but their final allowed values are
between {\it 10.2 -- 13.6}\sol. This indicates that an enormous
variety of yields can be generated from a single model by choosing the
parameters for explosion energy and amount of fallback and
mixing. \citet{lc03} examine nucleosynthesis with a variety of
explosion parameters and fallback. Their yields are highly variable
for Fe peak and intermediate mass elements, depending on the energy of
the explosion and the amount of fallback. Their range of variation for
CNO and Si is smaller than ours mainly because their range of
explosion energies does not extend as low as ours. We cannot compare
directly to their results because their progenitors do not include a
realistic prescription for mixing during the evolution, and therefore
the extent of the various nuclear processed regions is
smaller. However, if we examine their 35\sol progenitor we can make
some approximate comparisons since the extent of the oxygen and
heavier core is similar. They show a range of a factor of five in
$^{28}$Si and about an order of magnitude in $^{32}$S, $^{40}$Ca, and
$^{56}$Fe. There is little variation of CNO because the minimum
explosion energy is 1.3 foe, which ejects all or nearly all of the CNO
mantle. We find similar results for explosion energies this high.

Since we are required by computational limitations to rely on 1D
explosion for population yields, the best we can hope for in the near
future is use a few 3-dimensional simulations provide some insight
into energy distribution, the energy/remnant mass relation and 
amount of mixing to constrain the size of the 1-dimensional grid.  
But recall that the current uncertainties in the explosion mechanism, 
even in 3-dimensions, prevent such results from being the final yields 
of stars.  It is likely that we will have to live with uncertainties in 
the yields for some time yet, but understanding these uncertainties 
should help in understanding the current data.

\subsection{O-rich Supernova Remnants}

An incidental point of interest of this study concerns O-rich
supernova remnants. As a class these remnant have very large masses of
O dominated ejecta and comparatively small enrichments in Si/S and Fe
elements. In particular, N132D in the Large Magellanic Cloud (LMC) and
1E 0102.2-7219 in the SMC show no evidence for the products of oxygen
burning, either hydrostatic or explosive \citep{bl00}. In addition,
\citet{z03} present an analysis of two dim supernovae, SN 1997D and SN
1997br, which have small explosion energies ($<$1 foe) and very small
masses of $^{56}$Ni($< 10^{-2}$\sol). Taken together, these
observations are most simply explained by the explosion of a massive
star that does not eject material processed by oxygen burning or later
stages.

We see from these results that for a 23 \sol progenitor, explosion
energies of 1.2 and 1.5 foe for the piston and thermal bomb models,
respectively, are needed to produce a significant amount of
$^{56}$Ni. Small amounts, a few $\times 10^3$ \sol, can be explained
by 3D effects and mixing, even in weak explosions, but there is a very
large discrepancy between the energy necessary for significant
$^{56}$Ni production and the energies predicted for stars above 20
\sol by \citet{fk01}. We thus expect most stars between $\sim$ 20 \sol
and the lower mass limit for Wolf-Rayet stars (30 \sol?) to eject very
little Si and Fe peak, which represents a significant population of
objects that both explode as dim supernovae and produce O-rich
remnants.

\acknowledgements This work was carried out in part under the auspices
of the National Nuclear Security Administration of the U.S. Department
of Energy at Los Alamos National Laboratory and supported by Contract
No. DE-AC52-06NA25396, by a DOE SciDAC grant DE-FC02-01ER41176, an
NNSA ASC grant, and a subcontract to the ASCI FLASH Center at the
University of Chicago.

\clearpage

\begin{deluxetable}{lccc}
\tablewidth{0pt}
\tablecaption{Explosion Simulations\label{tab1}}
\tablehead{
  \colhead{Simulation}
& \colhead{type}
& \colhead{Energy}
& \colhead{M$_{\rm Rem}$}  \\

  \colhead{}
& \colhead{}
& \colhead{$10^{51}$\,erg}
& \colhead{(\sol)}

}
\startdata
23e-0.8 & thermal bomb - 20\,ms & 0.8 & 5.7 \\
23e-1.1 & thermal bomb - 20\,ms & 1.1 & 4.2 \\
23e-1.2 & thermal bomb - 20\,ms & 1.2 & 4.1 \\
23e-1.4 & thermal bomb - 20\,ms & 1.4 & 3.7 \\
23e-1.5 & thermal bomb - 20\,ms & 1.5 & 3.2 \\
23e-2.0 & thermal bomb - 20\,ms & 2.0 & 2.6 \\
23e-0.2-0.8 & thermal bomb - 200\,ms & 0.8 & 3.5 \\
23e-0.2-1.5 & thermal bomb - 200\,ms & 1.5 & 2.6 \\
23e-0.7-0.8 & thermal bomb - 700\,ms & 0.8 & 3.2 \\
23e-0.7-1.5 & thermal bomb - 700\,ms & 1.5 & 2.3 \\
23p-0.9 & piston       & 0.9 & 5.0 \\
23p-1.2 & piston       & 1.2 & 3.2 \\
23p-1.6 & piston       & 1.6 & 2.4 \\
\enddata
\end{deluxetable}

\begin{deluxetable}{lcccccc}
\tablewidth{0pt}
\tablecaption{Detailed Yields (\sol) Before Radioactive Decay For Thermal Bomb Simulations\label{tab2}}
\tablehead{
  \colhead{Species}
& \colhead{23e-0.8}
& \colhead{23e-1.1}
& \colhead{23e-1.2}
& \colhead{23e-1.4}
& \colhead{23e-1.5}
& \colhead{23e-2.0} \\ }
\startdata
$^{ 1}$H  & 1.46E+00 & 1.46E+00 & 1.46E+00 & 1.46E+00 & 1.46E+00 & 1.46E+00 \\
$^{ 2}$H  & 3.86E-16 & 4.46E-16 & 4.39E-16 & 4.78E-16 & 4.76E-16 & 5.96E-16 \\ 
$^{ 3}$H  & 2.39E-23 & 1.58E-16 & 1.58E-16 & 1.58E-16 & 1.57E-16 & 1.55E-16 \\ 
$^{ 3}$He & 6.60E-05 & 6.60E-05 & 6.60E-05 & 6.60E-05 & 6.60E-05 & 6.60E-05 \\
$^{ 4}$He & 1.13E+00 & 1.13E+00 & 1.13E+00 & 1.13E+00 & 1.13E+00 & 1.13E+00 \\ 
$^{ 6}$Li & 2.01E-18 & 2.35E-18 & 2.29E-18 & 2.51E-18 & 2.50E-18 & 3.22E-18 \\ 
$^{ 7}$Li & 2.89E-10 & 2.89E-10 & 2.89E-10 & 2.89E-10 & 2.89E-10 & 2.89E-10 \\ 
$^{ 7}$Be & 1.28E-10 & 1.28E-10 & 1.28E-10 & 1.28E-10 & 1.28E-10 & 1.28E-10 \\
\enddata
\end{deluxetable}

\begin{deluxetable}{lccc}
\tablewidth{0pt}
\tablecaption{Detailed Yields (\sol) Before Radioactive Decay For Piston Simulations\label{tab3}}
\tablehead{
  \colhead{Species}
& \colhead{23p-0.9}
& \colhead{23p-1.2}
& \colhead{23p-1.6} \\ }
\startdata
$^{ 1}$H  & 1.46E+00 & 1.46E+00 & 1.46E+00 \\ 
$^{ 2}$H  & 1.75E-16 & 1.67E-16 & 1.89E-16 \\
$^{ 3}$H  & 6.77E-18 & 1.77E-16 & 1.75E-16 \\
$^{ 3}$He & 6.60E-05 & 6.60E-05 & 6.60E-05 \\ 
$^{ 4}$He & 1.13E+00 & 1.13E+00 & 1.15E+00 \\ 
$^{ 6}$Li & 1.26E-18 & 4.24E-18 & 7.05E-18 \\
$^{ 7}$Li & 2.89E-10 & 2.89E-10 & 2.89E-10 \\
$^{ 7}$Be & 1.28E-10 & 1.28E-10 & 1.28E-10 \\
\enddata
\end{deluxetable}

\begin{deluxetable}{lcccc}
\tablewidth{0pt}
\tablecaption{Detailed Yields (\sol) Before Radioactive Decay For Delayed Thermal Bombs\label{tab4}}
\tablehead{
  \colhead{Species}
& \colhead{d0.2s0.8}
& \colhead{d0.2s1.5}
& \colhead{d0.7s0.8}
& \colhead{d0.7s1.5} \\ }
\startdata
$^{ 1}$H  & 1.47E+00 & 1.47E+00 & 1.47E+00 & 1.47E+00 \\  
$^{ 2}$H  & 3.03E-16 & 4.79E-16 & 3.86E-16 & 5.27E-16 \\  
$^{ 3}$H  & 1.63E-16 & 1.66E-16 & 1.52E-16 & 1.66E-16 \\  
$^{ 3}$He & 6.63E-05 & 6.63E-05 & 6.63E-05 & 6.63E-05 \\  
$^{ 4}$He & 1.13E+00 & 1.12E+00 & 1.13E+00 & 1.13E+00 \\  
$^{ 6}$Li & 1.45E-18 & 2.45E-18 & 1.79E-18 & 2.53E-18 \\  
$^{ 7}$Li & 2.90E-10 & 2.90E-10 & 2.90E-10 & 2.90E-10 \\  
$^{ 7}$Be & 1.29E-10 & 1.29E-10 & 1.29E-10 & 1.29E-10 \\
\enddata
\end{deluxetable}

\end{document}